\documentclass[longauth]{aa}
\usepackage{url,hyperref}
\usepackage{footmisc}
\usepackage{epstopdf}
\usepackage[varg]{txfonts}
\usepackage[usenames,dvipsnames]{color}
\usepackage[percent]{overpic}
\usepackage{booktabs}
\usepackage{relsize}

 \newcommand\doublerule{\toprule\toprule}

\makeatletter
\renewcommand*{\@fnsymbol}[1]{\ifcase#1\or*\or$\dagger$\or$\ddagger$\or**\or$\dagger\dagger$\or$\ddagger\ddagger$\fi}
\makeatother

\usepackage{amstext}

\begin{document}


\title{Particle transport within the pulsar wind nebula HESS\,J1825\,$-$137\thanks{All sky maps (FITS files) and spectra are available at the CDS via anonymous ftp to \url{cdsarc.u-strasbg.fr} (130.79.128.5) or via \url{http://cdsweb.u-strasbg.fr/cgi-bin/qcat?J/A+A/}.}}

\author{H.E.S.S. Collaboration
\and H.~Abdalla \inst{\ref{NWU}}
\and F.~Aharonian \inst{\ref{MPIK},\ref{DIAS},\ref{NASRA}}
\and F.~Ait~Benkhali \inst{\ref{MPIK}}
\and E.O.~Ang\"uner \inst{\ref{CPPM}}
\and M.~Arakawa \inst{\ref{Rikkyo}}
\and C.~Arcaro \inst{\ref{NWU}}
\and C.~Armand \inst{\ref{LAPP}}
\and M.~Arrieta \inst{\ref{LUTH}}
\and M.~Backes \inst{\ref{UNAM},\ref{NWU}}
\and M.~Barnard \inst{\ref{NWU}}
\and Y.~Becherini \inst{\ref{Linnaeus}}
\and J.~Becker~Tjus \inst{\ref{RUB}}
\and D.~Berge \inst{\ref{DESY}}
\and K.~Bernl\"ohr \inst{\ref{MPIK}}
\and R.~Blackwell \inst{\ref{Adelaide}}
\and M.~B\"ottcher \inst{\ref{NWU}}
\and C.~Boisson \inst{\ref{LUTH}}
\and J.~Bolmont \inst{\ref{LPNHE}}
\and S.~Bonnefoy \inst{\ref{DESY}}
\and P.~Bordas \inst{\ref{MPIK}}
\and J.~Bregeon \inst{\ref{LUPM}}
\and F.~Brun \inst{\ref{CENB}}
\and P.~Brun \inst{\ref{IRFU}}
\and M.~Bryan \inst{\ref{GRAPPA}}
\and M.~B\"{u}chele \inst{\ref{ECAP}}
\and T.~Bulik \inst{\ref{UWarsaw}}
\and T.~Bylund \inst{\ref{Linnaeus}}
\and M.~Capasso \inst{\ref{IAAT}}
\and S.~Caroff \inst{\ref{LLR}}\protect\footnotemark[1]
\and A.~Carosi \inst{\ref{LAPP}}
\and S.~Casanova \inst{\ref{IFJPAN},\ref{MPIK}}
\and M.~Cerruti \inst{\ref{LPNHE}}
\and N.~Chakraborty \inst{\ref{MPIK}}
\and T.~Chand \inst{\ref{NWU}}
\and S.~Chandra \inst{\ref{NWU}}
\and R.C.G.~Chaves \inst{\ref{LUPM},\ref{CurieChaves}}
\and A.~Chen \inst{\ref{WITS}}
\and S.~Colafrancesco \inst{\ref{WITS}} \protect\footnotemark[2] 
\and B.~Condon \inst{\ref{CENB}}
\and I.D.~Davids \inst{\ref{UNAM}}
\and C.~Deil \inst{\ref{MPIK}}
\and J.~Devin \inst{\ref{LUPM}}
\and P.~deWilt \inst{\ref{Adelaide}}
\and L.~Dirson \inst{\ref{HH}}
\and A.~Djannati-Ata\"i \inst{\ref{APC}}
\and A.~Dmytriiev \inst{\ref{LUTH}}
\and A.~Donath \inst{\ref{MPIK}}
\and V.~Doroshenko \inst{\ref{IAAT}}
\and L.O'C.~Drury \inst{\ref{DIAS}}
\and J.~Dyks \inst{\ref{NCAC}}
\and K.~Egberts \inst{\ref{UP}}
\and G.~Emery \inst{\ref{LPNHE}}
\and J.-P.~Ernenwein \inst{\ref{CPPM}}
\and S.~Eschbach \inst{\ref{ECAP}}
\and S.~Fegan \inst{\ref{LLR}}
\and A.~Fiasson \inst{\ref{LAPP}}
\and G.~Fontaine \inst{\ref{LLR}}
\and S.~Funk \inst{\ref{ECAP}}
\and M.~F\"u{\ss}ling \inst{\ref{DESY}}
\and S.~Gabici \inst{\ref{APC}}
\and Y.A.~Gallant \inst{\ref{LUPM}}
\and F.~Gat{\'e} \inst{\ref{LAPP}}
\and G.~Giavitto \inst{\ref{DESY}}
\and D.~Glawion \inst{\ref{LSW}}
\and J.F.~Glicenstein \inst{\ref{IRFU}}
\and D.~Gottschall \inst{\ref{IAAT}}
\and M.-H.~Grondin \inst{\ref{CENB}}
\and J.~Hahn \inst{\ref{MPIK}}
\and M.~Haupt \inst{\ref{DESY}}
\and G.~Heinzelmann \inst{\ref{HH}}
\and G.~Henri \inst{\ref{Grenoble}}
\and G.~Hermann \inst{\ref{MPIK}}
\and J.A.~Hinton \inst{\ref{MPIK}}
\and W.~Hofmann \inst{\ref{MPIK}}
\and C.~Hoischen \inst{\ref{UP}}
\and T.~L.~Holch \inst{\ref{HUB}}
\and M.~Holler \inst{\ref{LFUI}}
\and D.~Horns \inst{\ref{HH}}
\and D.~Huber \inst{\ref{LFUI}}
\and H.~Iwasaki \inst{\ref{Rikkyo}}
\and A.~Jacholkowska \inst{\ref{LPNHE}} \protect\footnotemark[2] 
\and M.~Jamrozy \inst{\ref{UJK}}
\and D.~Jankowsky \inst{\ref{ECAP}}
\and F.~Jankowsky \inst{\ref{LSW}}
\and L.~Jouvin \inst{\ref{APC}}
\and I.~Jung-Richardt \inst{\ref{ECAP}}
\and M.A.~Kastendieck \inst{\ref{HH}}
\and K.~Katarzy{\'n}ski \inst{\ref{NCUT}}
\and M.~Katsuragawa \inst{\ref{KAVLI}}
\and U.~Katz \inst{\ref{ECAP}}
\and D.~Kerszberg \inst{\ref{LPNHE}}
\and D.~Khangulyan \inst{\ref{Rikkyo}}
\and B.~Kh\'elifi \inst{\ref{APC}}
\and J.~King \inst{\ref{MPIK}}
\and S.~Klepser \inst{\ref{DESY}}
\and W.~Klu\'{z}niak \inst{\ref{NCAC}}
\and Nu.~Komin \inst{\ref{WITS}}
\and K.~Kosack \inst{\ref{IRFU}}
\and M.~Kraus \inst{\ref{ECAP}}
\and G.~Lamanna \inst{\ref{LAPP}}
\and J.~Lau \inst{\ref{Adelaide}}
\and J.~Lefaucheur \inst{\ref{LUTH}}
\and A.~Lemi\`ere \inst{\ref{APC}}
\and M.~Lemoine-Goumard \inst{\ref{CENB}}
\and J.-P.~Lenain \inst{\ref{LPNHE}}
\and E.~Leser \inst{\ref{UP}}
\and T.~Lohse \inst{\ref{HUB}}
\and R.~L\'opez-Coto \inst{\ref{MPIK}}
\and I.~Lypova \inst{\ref{DESY}}
\and D.~Malyshev \inst{\ref{IAAT}}
\and V.~Marandon \inst{\ref{MPIK}}
\and A.~Marcowith \inst{\ref{LUPM}}
\and C.~Mariaud \inst{\ref{LLR}}
\and G.~Mart\'i-Devesa \inst{\ref{LFUI}}
\and R.~Marx \inst{\ref{MPIK}}
\and G.~Maurin \inst{\ref{LAPP}}
\and P.J.~Meintjes \inst{\ref{UFS}}
\and A.M.W.~Mitchell \inst{\ref{MPIK},\ref{MitchellNowAt}}\protect\footnotemark[1]
\and R.~Moderski \inst{\ref{NCAC}}
\and M.~Mohamed \inst{\ref{LSW}}
\and L.~Mohrmann \inst{\ref{ECAP}}
\and C.~Moore \inst{\ref{Leicester}}
\and E.~Moulin \inst{\ref{IRFU}}
\and T.~Murach \inst{\ref{DESY}}
\and S.~Nakashima  \inst{\ref{RIKKEN}}
\and M.~de~Naurois \inst{\ref{LLR}}
\and H.~Ndiyavala  \inst{\ref{NWU}}
\and F.~Niederwanger \inst{\ref{LFUI}}
\and J.~Niemiec \inst{\ref{IFJPAN}}
\and L.~Oakes \inst{\ref{HUB}}
\and P.~O'Brien \inst{\ref{Leicester}}
\and H.~Odaka \inst{\ref{Tokyo}}
\and S.~Ohm \inst{\ref{DESY}}
\and M.~Ostrowski \inst{\ref{UJK}}
\and I.~Oya \inst{\ref{DESY}}
\and M.~Panter \inst{\ref{MPIK}}
\and R.D.~Parsons \inst{\ref{MPIK}}\protect\footnotemark[1]
\and C.~Perennes \inst{\ref{LPNHE}}
\and P.-O.~Petrucci \inst{\ref{Grenoble}}
\and B.~Peyaud \inst{\ref{IRFU}}
\and Q.~Piel \inst{\ref{LAPP}}
\and S.~Pita \inst{\ref{APC}}
\and V.~Poireau \inst{\ref{LAPP}}
\and A.~Priyana~Noel \inst{\ref{UJK}}
\and D.A.~Prokhorov \inst{\ref{WITS}}
\and H.~Prokoph \inst{\ref{DESY}}
\and G.~P\"uhlhofer \inst{\ref{IAAT}}
\and M.~Punch \inst{\ref{APC},\ref{Linnaeus}}
\and A.~Quirrenbach \inst{\ref{LSW}}
\and S.~Raab \inst{\ref{ECAP}}
\and R.~Rauth \inst{\ref{LFUI}}
\and A.~Reimer \inst{\ref{LFUI}}
\and O.~Reimer \inst{\ref{LFUI}}
\and M.~Renaud \inst{\ref{LUPM}}
\and F.~Rieger \inst{\ref{MPIK},\ref{FellowRieger}}
\and L.~Rinchiuso \inst{\ref{IRFU}}
\and C.~Romoli \inst{\ref{MPIK}}
\and G.~Rowell \inst{\ref{Adelaide}}
\and B.~Rudak \inst{\ref{NCAC}}
\and E.~Ruiz-Velasco \inst{\ref{MPIK}}
\and V.~Sahakian \inst{\ref{YPI},\ref{NASRA}}
\and S.~Saito \inst{\ref{Rikkyo}}
\and D.A.~Sanchez \inst{\ref{LAPP}}
\and A.~Santangelo \inst{\ref{IAAT}}
\and M.~Sasaki \inst{\ref{ECAP}}
\and R.~Schlickeiser \inst{\ref{RUB}}
\and F.~Sch\"ussler \inst{\ref{IRFU}}
\and A.~Schulz \inst{\ref{DESY}}
\and H.~Schutte \inst{\ref{NWU}}
\and U.~Schwanke \inst{\ref{HUB}}
\and S.~Schwemmer \inst{\ref{LSW}}
\and M.~Seglar-Arroyo \inst{\ref{IRFU}}
\and M.~Senniappan \inst{\ref{Linnaeus}}
\and A.S.~Seyffert \inst{\ref{NWU}}
\and N.~Shafi \inst{\ref{WITS}}
\and I.~Shilon \inst{\ref{ECAP}}
\and K.~Shiningayamwe \inst{\ref{UNAM}}
\and R.~Simoni \inst{\ref{GRAPPA}}
\and A.~Sinha \inst{\ref{APC}}
\and H.~Sol \inst{\ref{LUTH}}
\and A.~Specovius \inst{\ref{ECAP}}
\and M.~Spir-Jacob \inst{\ref{APC}}
\and {\L.}~Stawarz \inst{\ref{UJK}}
\and R.~Steenkamp \inst{\ref{UNAM}}
\and C.~Stegmann \inst{\ref{UP},\ref{DESY}}
\and C.~Steppa \inst{\ref{UP}}
\and T.~Takahashi  \inst{\ref{KAVLI}}
\and J.-P.~Tavernet \inst{\ref{LPNHE}}
\and T.~Tavernier \inst{\ref{IRFU}}
\and A.M.~Taylor \inst{\ref{DESY}}
\and R.~Terrier \inst{\ref{APC}}
\and L.~Tibaldo \inst{\ref{MPIK}}
\and D.~Tiziani \inst{\ref{ECAP}}
\and M.~Tluczykont \inst{\ref{HH}}
\and C.~Trichard \inst{\ref{LLR}}
\and M.~Tsirou \inst{\ref{LUPM}}
\and N.~Tsuji \inst{\ref{Rikkyo}}
\and R.~Tuffs \inst{\ref{MPIK}}
\and Y.~Uchiyama \inst{\ref{Rikkyo}}
\and D.J.~van~der~Walt \inst{\ref{NWU}}
\and C.~van~Eldik \inst{\ref{ECAP}}
\and C.~van~Rensburg \inst{\ref{NWU}}
\and B.~van~Soelen \inst{\ref{UFS}}
\and G.~Vasileiadis \inst{\ref{LUPM}}
\and J.~Veh \inst{\ref{ECAP}}
\and C.~Venter \inst{\ref{NWU}}
\and P.~Vincent \inst{\ref{LPNHE}}
\and J.~Vink \inst{\ref{GRAPPA}}
\and F.~Voisin \inst{\ref{Adelaide}}
\and H.J.~V\"olk \inst{\ref{MPIK}}
\and T.~Vuillaume \inst{\ref{LAPP}}
\and Z.~Wadiasingh \inst{\ref{NWU}}
\and S.J.~Wagner \inst{\ref{LSW}}
\and R.M.~Wagner \inst{\ref{OKC}}
\and R.~White \inst{\ref{MPIK}}
\and A.~Wierzcholska \inst{\ref{IFJPAN}}
\and R.~Yang \inst{\ref{MPIK}}
\and H.~Yoneda \inst{\ref{KAVLI}}
\and D.~Zaborov \inst{\ref{LLR}}
\and M.~Zacharias \inst{\ref{NWU}}
\and R.~Zanin \inst{\ref{MPIK}}
\and A.A.~Zdziarski \inst{\ref{NCAC}}
\and A.~Zech \inst{\ref{LUTH}}
\and F.~Zefi \inst{\ref{LLR}}
\and A.~Ziegler \inst{\ref{ECAP}}
\and J.~Zorn \inst{\ref{MPIK}}
\and N.~\.Zywucka \inst{\ref{UJK}}
}

\institute{
Centre for Space Research, North-West University, Potchefstroom 2520, South Africa \label{NWU} \and 
Universit\"at Hamburg, Institut f\"ur Experimentalphysik, Luruper Chaussee 149, D 22761 Hamburg, Germany \label{HH} \and 
Max-Planck-Institut f\"ur Kernphysik, P.O. Box 103980, D 69029 Heidelberg, Germany \label{MPIK} \and 
Dublin Institute for Advanced Studies, 31 Fitzwilliam Place, Dublin 2, Ireland \label{DIAS} \and 
National Academy of Sciences of the Republic of Armenia,  Marshall Baghramian Avenue, 24, 0019 Yerevan, Republic of Armenia  \label{NASRA} \and
Yerevan Physics Institute, 2 Alikhanian Brothers St., 375036 Yerevan, Armenia \label{YPI} \and
Institut f\"ur Physik, Humboldt-Universit\"at zu Berlin, Newtonstr. 15, D 12489 Berlin, Germany \label{HUB} \and
University of Namibia, Department of Physics, Private Bag 13301, Windhoek, Namibia \label{UNAM} \and
GRAPPA, Anton Pannekoek Institute for Astronomy, University of Amsterdam,  Science Park 904, 1098 XH Amsterdam, The Netherlands \label{GRAPPA} \and
Department of Physics and Electrical Engineering, Linnaeus University,  351 95 V\"axj\"o, Sweden \label{Linnaeus} \and
Institut f\"ur Theoretische Physik, Lehrstuhl IV: Weltraum und Astrophysik, Ruhr-Universit\"at Bochum, D 44780 Bochum, Germany \label{RUB} \and
Institut f\"ur Astro- und Teilchenphysik, Leopold-Franzens-Universit\"at Innsbruck, A-6020 Innsbruck, Austria \label{LFUI} \and
School of Physical Sciences, University of Adelaide, Adelaide 5005, Australia \label{Adelaide} \and
LUTH, Observatoire de Paris, PSL Research University, CNRS, Universit\'e Paris Diderot, 5 Place Jules Janssen, 92190 Meudon, France \label{LUTH} \and
Sorbonne Universit\'e, Universit\'e Paris Diderot, Sorbonne Paris Cit\'e, CNRS/IN2P3, Laboratoire de Physique Nucl\'eaire et de Hautes Energies, LPNHE, 4 Place Jussieu, F-75252 Paris, France \label{LPNHE} \and
Laboratoire Univers et Particules de Montpellier, Universit\'e Montpellier, CNRS/IN2P3,  CC 72, Place Eug\`ene Bataillon, F-34095 Montpellier Cedex 5, France \label{LUPM} \and
IRFU, CEA, Universit\'e Paris-Saclay, F-91191 Gif-sur-Yvette, France \label{IRFU} \and
Astronomical Observatory, The University of Warsaw, Al. Ujazdowskie 4, 00-478 Warsaw, Poland \label{UWarsaw} \and
Aix Marseille Universit\'e, CNRS/IN2P3, CPPM, Marseille, France \label{CPPM} \and
Instytut Fizyki J\c{a}drowej PAN, ul. Radzikowskiego 152, 31-342 Krak{\'o}w, Poland \label{IFJPAN} \and
Funded by EU FP7 Marie Curie, grant agreement No. PIEF-GA-2012-332350 \label{CurieChaves}  \and
School of Physics, University of the Witwatersrand, 1 Jan Smuts Avenue, Braamfontein, Johannesburg, 2050 South Africa \label{WITS} \and
Laboratoire d'Annecy de Physique des Particules, Univ. Grenoble Alpes, Univ. Savoie Mont Blanc, CNRS, LAPP, 74000 Annecy, France \label{LAPP} \and
Landessternwarte, Universit\"at Heidelberg, K\"onigstuhl, D 69117 Heidelberg, Germany \label{LSW} \and
Universit\'e Bordeaux, CNRS/IN2P3, Centre d'\'Etudes Nucl\'eaires de Bordeaux Gradignan, 33175 Gradignan, France \label{CENB} \and
Oskar Klein Centre, Department of Physics, Stockholm University, Albanova University Center, SE-10691 Stockholm, Sweden \label{OKC} \and
Institut f\"ur Astronomie und Astrophysik, Universit\"at T\"ubingen, Sand 1, D 72076 T\"ubingen, Germany \label{IAAT} \and
Laboratoire Leprince-Ringuet, Ecole Polytechnique, CNRS/IN2P3, F-91128 Palaiseau, France \label{LLR} \and
APC, AstroParticule et Cosmologie, Universit\'{e} Paris Diderot, CNRS/IN2P3, CEA/Irfu, Observatoire de Paris, Sorbonne Paris Cit\'{e}, 10, rue Alice Domon et L\'{e}onie Duquet, 75205 Paris Cedex 13, France \label{APC} \and
Univ. Grenoble Alpes, CNRS, IPAG, F-38000 Grenoble, France \label{Grenoble} \and
Department of Physics and Astronomy, The University of Leicester, University Road, Leicester, LE1 7RH, United Kingdom \label{Leicester} \and
Nicolaus Copernicus Astronomical Center, Polish Academy of Sciences, ul. Bartycka 18, 00-716 Warsaw, Poland \label{NCAC} \and
Institut f\"ur Physik und Astronomie, Universit\"at Potsdam,  Karl-Liebknecht-Strasse 24/25, D 14476 Potsdam, Germany \label{UP} \and
Friedrich-Alexander-Universit\"at Erlangen-N\"urnberg, Erlangen Centre for Astroparticle Physics, Erwin-Rommel-Str. 1, D 91058 Erlangen, Germany \label{ECAP} \and
DESY, D-15738 Zeuthen, Germany \label{DESY} \and
Obserwatorium Astronomiczne, Uniwersytet Jagiello{\'n}ski, ul. Orla 171, 30-244 Krak{\'o}w, Poland \label{UJK} \and
Centre for Astronomy, Faculty of Physics, Astronomy and Informatics, Nicolaus Copernicus University,  Grudziadzka 5, 87-100 Torun, Poland \label{NCUT} \and
Department of Physics, University of the Free State,  PO Box 339, Bloemfontein 9300, South Africa \label{UFS} \and
Heisenberg Fellow (DFG), ITA Universit\"at Heidelberg, Germany \label{FellowRieger} \and
Department of Physics, Rikkyo University, 3-34-1 Nishi-Ikebukuro, Toshima-ku, Tokyo 171-8501, Japan \label{Rikkyo} \and
Kavli Institute for the Physics and Mathematics of the Universe (Kavli IPMU), The University of Tokyo Institutes for Advanced Study (UTIAS), The University of Tokyo, 5-1-5 Kashiwa-no-Ha, Kashiwa City, Chiba, 277-8583, Japan \label{KAVLI} \and
Department of Physics, The University of Tokyo, 7-3-1 Hongo, Bunkyo-ku, Tokyo 113-0033, Japan \label{Tokyo} \and
RIKEN, 2-1 Hirosawa, Wako, Saitama 351-0198, Japan \label{RIKKEN} \and
Now at The School of Physics, The University of New South Wales, Sydney, 2052, Australia \label{MaxtedNowAt} \and
Now at Instituto de F\'{i}sica de S\~{a}o Carlos, Universidade de S\~{a}o Paulo, Av. Trabalhador S\~{a}o-carlense, 400 - CEP 13566-590, S\~{a}o Carlos, SP, Brazil \label{VianaNowAt} \and
Now at Physik Institut, Universit\"{a}t Z\"{u}rich, Winterthurerstrasse 190, CH-8057 Z\"{u}rich, Switzerland \label{MitchellNowAt}
}

\offprints{H.E.S.S.~collaboration:
\protect\\\email{\href{mailto:contact.hess@hess-experiment.eu}{contact.hess@hess-experiment.eu}};
\protect\\\protect\footnotemark[1] Corresponding authors
\protect\\\protect\footnotemark[2] Deceased
}

\date{Received 28.09.2018 / Accepted 25.10.2018}

\abstract{\textit{Aims.} We present a detailed view of the pulsar wind nebula (PWN) HESS\,J1825--137. We aim to constrain the mechanisms dominating the particle transport within the nebula, accounting for its anomalously large size and spectral characteristics.\\ 
\textit{Methods.} The nebula was studied using a deep exposure from over 12 years of H.E.S.S. I operation, together with data from H.E.S.S. II that improve the low-energy sensitivity. Enhanced energy-dependent morphological and spatially resolved spectral analyses probe the very high energy (VHE, E$>$0.1\,TeV) $\gamma$-ray properties of the nebula. \\ 
\textit{Results.} The nebula emission is revealed to extend out to $1.5^\circ$ from the pulsar, $\sim$1.5 times farther than previously seen, making HESS\,J1825--137, with an intrinsic diameter of $\sim$100~pc, potentially the largest $\gamma$-ray PWN currently known. Characterising the strongly energy-dependent morphology of the nebula enables us to constrain the particle transport mechanisms. A dependence of the nebula extent with energy of $R\propto E^\alpha$ with $\alpha = -0.29\pm0.04_{\mathrm{stat}}\pm0.05_{\mathrm{sys}}$ disfavours a pure diffusion scenario for particle transport within the nebula.
The total $\gamma$-ray flux of the nebula above 1~TeV is found to be ($1.12 \pm 0.03_{\mathrm{stat}} \pm 0.25_{\mathrm{sys}} $) $\times 10^{-11}\mathrm{cm}^{-2}\mathrm{s}^{-1}$, corresponding to $\sim64\%$ of the flux of the Crab nebula.\\ 
\textit{Conclusions.} HESS\,J1825-137 is a PWN with clearly energy-dependent morphology at VHE $\gamma$-ray energies. This source is used as a laboratory to investigate particle transport within intermediate-age PWNe.  Based on deep observations of this highly spatially extended PWN, we produce a spectral map of the region that provides insights into the spectral variation within the nebula.  
}

\keywords{pulsar wind nebula; gamma-rays; PSR\,B1823--13; HESS\,J1825--137}

\maketitle

\makeatletter
\renewcommand*{\@fnsymbol}[1]{\ifcase#1\@arabic{#1}\fi}
\makeatother

\section{Introduction and multi-wavelength context}
\label{sec:intro}

Pulsar wind nebulae (PWNe) are one of the largest Galactic source classes at very high energy (VHE, E$>$0.1\,TeV) $\gamma$-ray energies, as recently demonstrated by the H.E.S.S. (High Energy Stereoscopic System) Galactic Plane Survey (HGPS), \cite{Abdalla18hgps}. The characteristics of the known VHE $\gamma$-ray PWNe were further explored in an associated PWNe population study, finding that these PWNe are associated with pulsars with a high spin-down luminosity ($\dot{E} > 10^{35}\mathrm{erg~s}^{-1}$) \citep{Abdalla17pwnpop}. 

Following a supernova explosion, PWNe may form around the pulsar PSR remnant of the progenitor star; the wind expands into the region that has been evacuated by the supernova ejecta.
Pulsars have a strong induced electric field on their surface and generate a plasma of electrons and positrons through pair production in the magnetosphere. (``Electrons'' here refers to both electrons and positrons.) This produces highly relativistic winds (with a frozen-in magnetic field) of electron-positron pairs that stream away from the pulsar region. Within this wind-dominated region, a significant proportion of the Poynting flux of the magnetic dipole radiation is converted into particle kinetic energy \citep{Rees74}. The mechanisms by which this occurs remain an active area of research \citep{Kirk03,Kirk09,Porth13}. As the ram pressure of the magnetised wind reduces to the pressure of the surrounding medium, a termination shock forms at the front of the expanding supernova \citep{Rees74,KennelCoroniti84A}. Charged particles can be further accelerated at this shock, for example, by diffusive shock acceleration, as long as they remain confined to the shock region \citep{ReynoldsChevalier84}. Relativistic particles downstream of the shock produce synchrotron and inverse Compton (IC) radiation that is detectable from radio through to X-rays and in the $\gamma$-ray region respectively. 
Pulsars are expected to be major contributors to the leptonic cosmic ray flux measured at Earth \citep{Shen70,Atoyan95}.

The expanding supernova shell slows down as it accumulates material in the surrounding interstellar medium (ISM); as the supernova enters the Sedov-Taylor phase, the shell comprises both a forward and reverse shock that eventually returns inwards \citep{Truelove99}.
The reverse shock formed in such systems may interact with the termination shock of the expanding PWN, leading to a crushing effect in some systems \citep{ReynoldsChevalier84}. As the pulsar continues to produce a wind-driven outflow, the system rebounds against crushing by the reverse shock with further PWN expansion. Complex reverberations follow, until over time, the system relaxes \citep{Gaensler06}. 
VHE $\gamma$-ray PWNe are often extended and offset from the associated pulsar because the supernova remnant (SNR) expands into an inhomogeneous medium, resulting in variation in reverse shock interaction times around the PWN \citep{deJager09,Gaensler06}. This evolutionary scenario is typically invoked to account for the wide variety of PWN morphology seen in older, evolved systems, with multiple examples of unique morphology known \citep{Lamanna12,Dalton12,KhelifiMSH05}. 
 When the SNR shell has dissipated, only the PWN remains detectable. Alternatively, PWN may interact with the supernova material and form a composite SNR-PWN system. 
 
PSR~B1823-13 (R.A.: $18^{h}26^{m}13.06^{s}$, Dec: $-13^\circ34'48.1''$, also known as PSR~J1826-1334), is a young pulsar (characteristic spin-down age $\tau = 2.14\times 10^{4}$~yr) with a spin-down power of $\dot{E}=2.8\times 10^{36}~\mathrm{erg~s}^{-1}$ and a period of $P=0.1015$~s, situated at a distance of approximately 4~kpc \citep{Manchester05}. It therefore has the characteristics of a Vela-like pulsar \citep{Aharonian06}.The distance estimate to PSR~B1823-13 has been revised multiple times in the range 3.6~kpc to 4.2~kpc since \citep{Funk06}. We adopt a 4~kpc distance for the nebula in this paper.

X-ray observations of PSR\,B1823--13 with ROSAT revealed a point source surrounded by a compact nebula of $\sim20''$ radius and a diffuse $\sim4'$ region of extended emission \citep{Finley96}. This X-ray detection of a synchrotron nebula was later confirmed by ASCA (Advanced Satellite for Cosmology and Astrophysics), who also found a point source with a surrounding diffuse nebula \citep{Sakurai01}.

XMM-Newton (X-ray Multi-Mirror Mission) observations confirmed the identification in X-rays of both a compact core and a diffuse component to the nebula, extending to 30'' and 5' , respectively. The morphology was found to be asymmetric, which was attributed to compression and distortion of the pulsar wind by an asymmetric reverse shock interaction \citep{Gaensler03}. Subsequent observations by the X-ray satellites Chandra and Suzaku used multiple observation positions to cover the extended nebula- They consistently identified a compact core and extended component to the nebula, with Suzaku measurements finding diffuse X-ray emission up to 15' (17pc) from the pulsar \citep{Pavlov08,Uchiyama09}. All X-ray observations consistently found a power-law spectrum, $\phi\propto E^{-\Gamma}$, with spectral index around $\Gamma\approx2$ for the extended diffuse component, and a somewhat harder power-law index of $\Gamma\approx1.3 - 1.6$ for the compact core. Additionally, \cite{Pavlov08} performed infrared (Spitzer GLIMPSE survey) and radio (NRAO (National Radio Astronomy Observatory) VLA (Very Large Array) Sky Survey) observations of the region at $8~\mu$m, $24~\mu$m and 1.4~GHz, but found no obvious counterparts to the nebula. 

Radio observations of PSR~B1823-13 with the VLA were used to derive its proper motion, finding a tangential velocity of $v_\perp =(443\pm46)~\mathrm{km~s}^{-1}$ at the nominal 4~kpc distance. This corresponds to a total distance of $8.2'$ travelled across the sky during the pulsar's characteristic age $\tau$ of 21~kyr, interestingly, in a direction almost perpendicular to that in which the nebula extends. \cite{Pavlov08} suggested that the large angle between the nebula extension and the pulsar's proper motion trajectory lends credence to an evolutionary scenario where the nebula shape was influenced at early times by a reverse-shock interaction. 
 
\begin{figure*}
\includegraphics[width=\columnwidth]{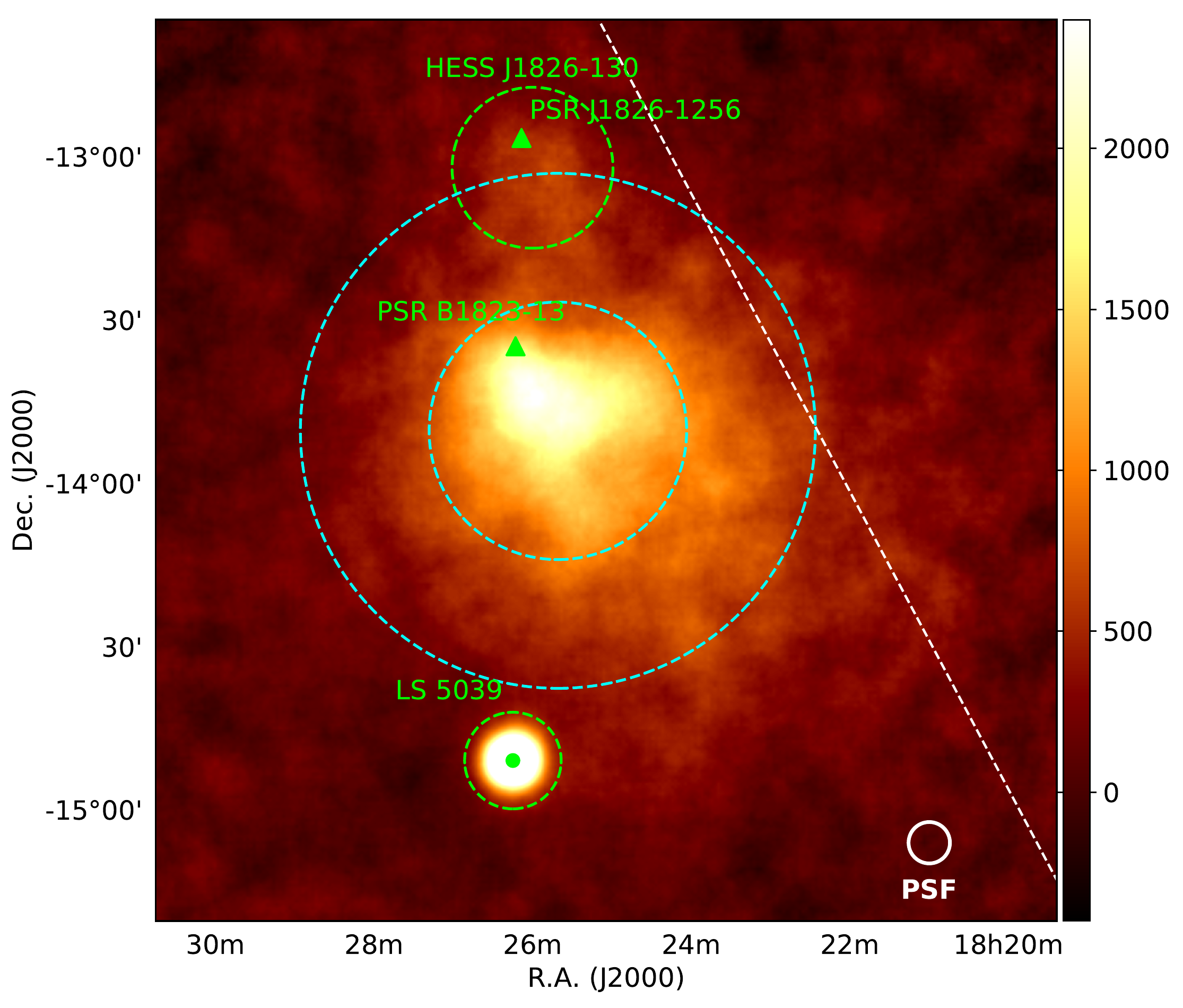}
\includegraphics[width=\columnwidth]{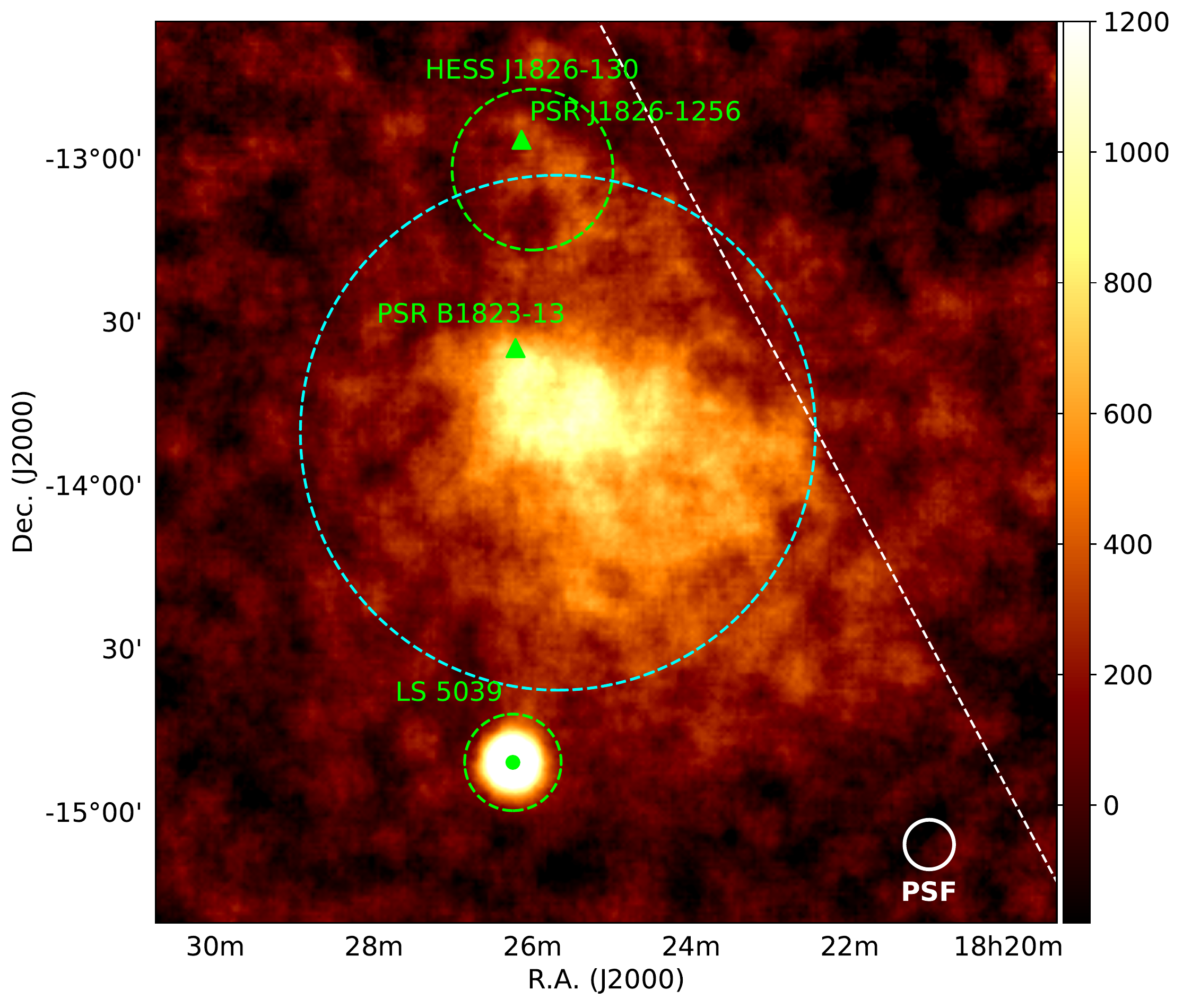}
\caption{Left: Excess count map of the nebula using analysis A, with the Galactic plane indicated by the dashed white line and the locations of two energetic pulsars in the region indicated by green triangles. The two spectral extraction regions used in figure \ref{fig:spectrum} are overlaid. The larger region with a  radius of $0.8^{\circ}$ slightly overlaps HESS\,J1826--130, (RA: $18^{h}26^{m}00^{s}$, Dec: $-13^\circ02'02''$), whose location and approximate extent is indicated by the green dashed circle, whereas the region with the smaller radius of $0.4^{\circ}$  encompasses the core emission and peak of the nebula. Both regions are centred on the best-fit position of the nebula as determined by \cite{Funk06}. Right: Excess count map of the nebula using analysis B, shown for comparison with the region of $0.8^{\circ}$ radius overlaid. The exposure times and telescope configurations for the two analyses are given in table \ref{tab:datasets}. A correlation radius of $0.07^\circ$ was used for both excess maps. }
\label{fig:sigmapAcorereg}
\end{figure*}

The H.E.S.S. experiment is an array of five imaging atmospheric Cherenkov telescopes (IACTs) and is situated in the Khomas highlands of Namibia \citep{Hinton04}. Comprised until 2012 of four $107\,\mathrm{m}^2$ mirror area IACTs (CT1-4), H.E.S.S. then entered a second phase of operation with the addition of a fifth $612\,\mathrm{m}^2$ mirror area IACT (CT5), which lowered the energy threshold \citep{Abdalla16mono,Holler15ICRC}. 

An associated VHE $\gamma$-ray nebula was discovered by H.E.S.S. in 2005 as part of the first H.E.S.S. Galactic plane survey (HGPS) \citep{Aharonian05pwn1825} and was given the identifier HESS\,J1825--137, with an extent in VHE $\gamma$-rays of $0.5^\circ$; this exceeds that of the X-ray nebula. It was found to be one of the brightest sources in the Milky Way at these very high energies \citep{Aharonian06survey}. A further detailed study made by H.E.S.S. in 2006 explored the nebula properties at TeV energies; in particular, the nebula was shown to exhibit a strongly energy-dependent morphology \citep{Funk06}. In spatially resolved spectra, the spectral index was seen to soften with increasing distance from the pulsar, which implies that the population of electrons in the nebula had travelled and cooled out to large distances. That the low-energy electrons at large distances from the pulsar produce the softest spectrum was interpreted to mean that these are the oldest electrons in the system.
Several mechanisms might account for this trend: radiative cooling of electrons as they propagate away from the pulsar and cause energy loss, energy-dependent transport mechanisms such as diffusion or advection, or a variation in the electron injection spectrum over time \citep{Funk06}. 

Two other prominent $\gamma$-ray sources lie in close proximity to HESS\,J1825--137: the binary system LS\,5039 to the south, and the unidentified hard-spectrum source HESS\,J1826--130 to the north. 
Previously thought to be part of the HESS\,J1825--137 complex, HESS\,J1826--130 has since been identified as a separate source in the HGPS \citep{Abdalla18hgps}.

In 2011, HESS\,J1825--137 was detected by Fermi-LAT (Fermi Large Area Telescope) in the energy range from 1--100 GeV. The best-fit rms extent with a Gaussian morphological model was  $\sigma = 0.56^\circ \pm 0.07^\circ$ \citep{Grondin11}. The emission is best described by a power law with spectral index $\Gamma = 1.38 \pm 0.12_{\mathrm{stat}} \pm 0.16_{\mathrm{sys}}$ and flux consistent with that reported by H.E.S.S., with the peak of the GeV emission spatially offset by $0.32^\circ$ from the TeV peak.

In the 2017 Fermi Galactic Extended Source Catalogue (FGES), a disc morphological model was adopted, with a somewhat larger radius of $\sigma = 1.05^\circ \pm 0.02_{\mathrm{stat}}^\circ \pm 0.25_{\mathrm{sys}}^\circ$. The spectrum obtained from 10~GeV to 2~TeV joined previous Fermi-LAT and H.E.S.S. results and confirmed that an IC peak occurs at around 100~GeV. A log-parabola spectral model was preferred over power-law models, with an index of $\Gamma = 1.3  \pm 0.1_{\mathrm{stat}}\pm 0.4_{\mathrm{sys}}$ and an integrated flux from 10\,GeV to 2\,TeV of $(19.59\pm 0.14_{\mathrm{stat}} \pm 0.22_{\mathrm{sys}}) \times 10^{-10}\mathrm{cm}^{-2}\mathrm{s}^{-1}$ with a curvature of $\beta = 0.27 \pm 0.05_{\mathrm{stat}} \pm 0.07_{\mathrm{sys}}$. The source FGES J1825.2-1359 alone in the FGES above 10~GeV has a log-parabola rather than power law as the best-fit spectral model \citep{Ackermann17}.

The detection of an extended faint radio counterpart to the nebula in extended VLA (EVLA) observations at 1.4 GHz was made in 2012 by \cite{Castelletti12}, who also found a nearby molecular cloud with a density of $\sim400~\mathrm{cm}^{-3}$. Investigation of interstellar gas properties (using molecular CO and atomic HI observations) revealed a dense molecular cloud north of the nebula, which is thought to help explain the apparent current asymmetric shape \citep{Lemiere05}. At all wavelengths, the nebula emission lies predominantly south of the pulsar. It is thought that at early times, the outward-moving SNR shell interacted with and was slowed down by the nearby molecular cloud, leading to comparatively rapid formation of a reverse shock on the northern side of the nebula. The return of this reverse shock pushed the nebula towards the southern side; this preference  has since remained in place, although the nebula has considerably broadened. 

Recent studies of the surrounding region by \cite{Voisin16} have claimed a potential association between the progenitor SNR of HESS\,J1825--137 and a H$\alpha$ ridge-like structure found by \cite{Stupar08}. Using millimetre observations with the Mopra and Nanten telescopes, \cite{Voisin16} identified six regions of interest that revealed several molecular gas clouds traced by a variety of spectral lines. 
Significant amounts of molecular gas were found towards HESS\,J1826--130 (in five of the six regions), with a sixth more southerly region of dense gas located within the $\gamma$-ray nebula.

Most recently, the HAWC (High Altitude Water Cherenkov) experiment detected highly significant emission coming from the HESS\,J1825--137 region as part of the HAWC Galactic Plane Survey, although due to its limited angular resolution, it was not possible to separately identify HESS\,J1825--137 and the neighbouring source HESS\,J1826--130. This led to the HAWC identifier 2HWC\,J1825--134 for the combined emission from the region \citep{Abeysekara17}.

\begin{table}
\centering
\begin{tabular}{ccccc}
Analysis & Telescopes & Exposure & Time Period & $\theta_z$\\
\doublerule
A & CT1-4 & 387 hours & 2004 - 2016  & $25.8^\circ$\\
B & CT1-5 & 136 hours & 2012 - 2016 & $23.2^\circ$
\end{tabular}
\caption{Summary of the data used in the two analyses, including the exposure and mean zenith angle $\theta_z$ of the observations. The data for analysis B were also used in analysis A, but without CT5. }
\label{tab:datasets}
\end{table}

\begin{figure*}
\begin{center}
 \includegraphics[trim=10mm 0mm 15mm 0mm,clip,width=\columnwidth]{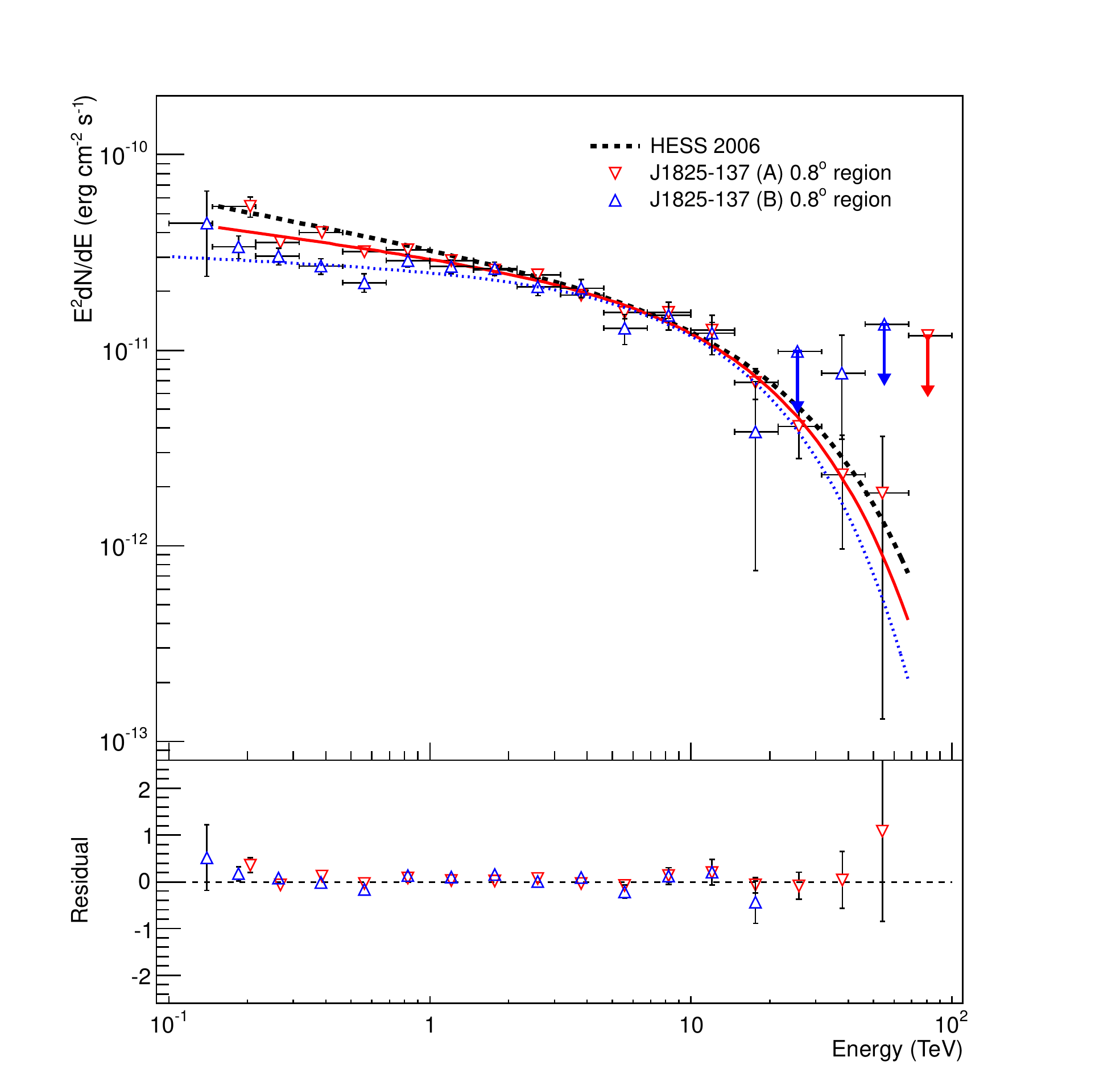}
 \includegraphics[trim=10mm 0mm 15mm 0mm,clip,width=\columnwidth]{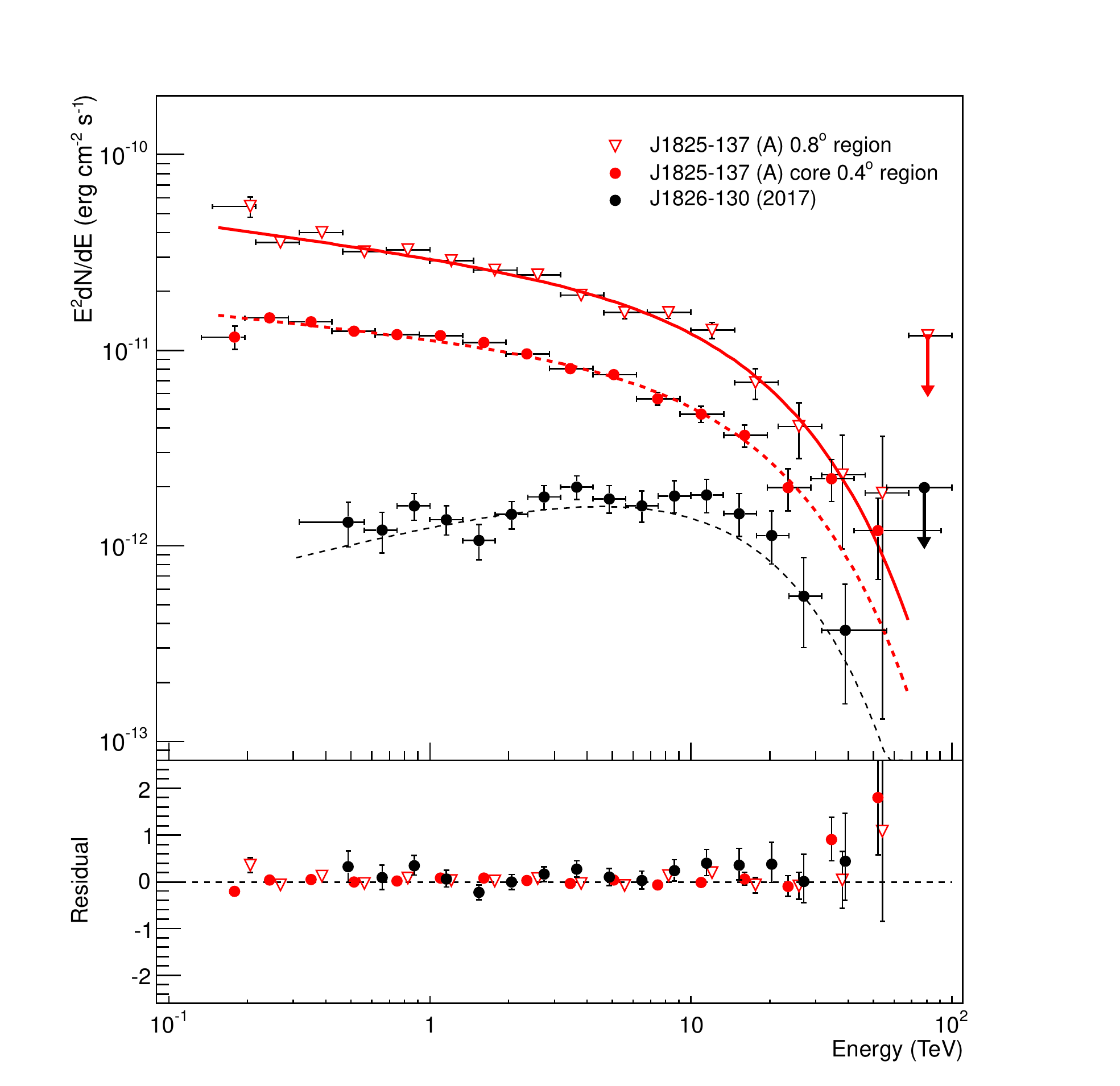}
\caption{Left: Spectra extracted from the region with a radius of $0.8^\circ$  shown in figure \ref{fig:sigmapAcorereg}, encompassing the majority of the nebula emission. The fit parameters given in table \ref{tab:spectralfits} for analyses A and B both agree well with the previously reported spectrum \citep{Funk06}. Right: Comparison of spectra extracted from the regions of $0.8^\circ$ and $0.4^\circ$ radii shown in figure \ref{fig:sigmapAcorereg} together with a spectrum of HESS\,J1826--130 \citep{Ang1826ICRC17}. All spectra are shown with a best-fit model of a power law with exponential cut-off.}
\label{fig:spectrum}
\end{center}
\end{figure*}

\section{H.E.S.S. data and analysis}
\label{sec:hessdata}

The H.E.S.S. dataset on the HESS\,J1825-137 region is considerably larger than that available to previous studies; together with improved analysis procedures, the sensitivity to weaker emission has been significantly enhanced by an increase in exposure of a factor of four with respect to \cite{Funk06}. 
For the purpose of this study, the H.E.S.S. data were analysed using standard event selection cuts in two different combinations, A and B, according to the configuration of telescopes that was used \citep{Aharonian06}. Analysis A comprises all data in which at least three of the four original H.E.S.S. telescopes (CT1-4) participated for a total exposure of $\sim400$ hours from data taken between 2004 and 2016 (see table \ref{tab:datasets}). Analysis B only includes data in which CT5 participated in the trigger, for a total exposure time of $\sim140$ hours.  All data were analysed in stereoscopic mode, that is, at least two telescopes contributed to each event. Consequently, most of the data comprising analysis B are also included in analysis A, but were analysed without CT5 for compatibility with the analysis of pre-2012 data. We note that $\sim226$ hours of the analysis A exposure stem from data post-2012; considerably more time than in analysis B, as CT5 was not always present in observations of this region. 
Analysis A has the advantage of deep exposure with good resolution, whilst analysis B, which always includes CT5, offers a lower energy threshold. The mean zenith angle of the observations in both analyses is given in table \ref{tab:datasets}. The energy threshold of H.E.S.S. II for a spectral analysis with analysis B is around 140\,GeV, whilst the threshold is around 200\,GeV for analysis A. These two thresholds are both much lower than the previous threshold of 270\,GeV \citep{Funk06}.  Unless otherwise specified, the results presented in this paper are from analysis A. 

A sensitive likelihood-based template fitting analysis was used for this study \citep{Parsons14}. All results from both analyses were cross-checked using an independent calibration and analysis chain \citep{deNaurois09}, and were found to be consistent within systematic errors. In the case of H.E.S.S. I, the contributions to systematic errors that affect the spectral analysis are given in \cite{Aharonian06}.

Two methods of background estimation were used in this analysis. For morphological studies and the generation of sky maps, the ring background method was applied, whilst for spectral analyses, the reflected background method was used. These methods and their comparative suitability for different types of analysis are discussed in \cite{BergeFunkHinton07}. Both methods designate a particular region within the field of view as the source ON region (the ON region may be called the ``region of interest'' in the case of potential sources)  and an independent region within the field of view as being suitable for background estimation (the OFF region).

\begin{table*}
\begin{center}
\begin{tabular}{ccccccc}
\hline
Analysis & Region &  $\phi$ Fit Model & $I_0$ & $\Gamma$  & Fit Parameters  & $\chi^2 / \mathrm{ndf}$ \\
\hline
& & $I_0 \left(\frac{E}{E_0}\right)^{-\Gamma}$ & $6.81\pm 0.07\pm 0.2$ & $2.28\pm 0.01\pm 0.02$ & - & 141/14  \\
A & $0.4^{\circ}$ & $I_0 \left(\frac{E}{E_0}\right)^{-\Gamma}\exp\left(-\frac{E}{E_c}\right)$ & $7.20\pm 0.09\pm 0.2$ & $2.13\pm 0.02\pm 0.03$ & $E_c = 19 \pm 3\pm 0.8$ TeV & 21/13 \\
& & $I_0 \left(\frac{E}{E_0}\right)^{-\Gamma + \beta\log (E/E_0)}$ & $7.4\pm 0.1\pm 0.1$ & $2.26\pm 0.01\pm 0.02$ & $\beta= 0.078 \pm 0.008\pm 0.01$ & 21/13 \\
\doublerule 
& & $I_0 \left(\frac{E}{E_0}\right)^{-\Gamma}$ & $17.9\pm 0.2\pm 0.4$ & $2.33\pm 0.01\pm 0.01$ & - & 134/15 \\
A & $0.8^{\circ}$ & $I_0 \left(\frac{E}{E_0}\right)^{-\Gamma}\exp\left(-\frac{E}{E_c}\right)$ & $18.8\pm 0.2\pm 0.3$ & $2.18\pm 0.02\pm 0.02$ & $E_c = 19\pm 3\pm 2$ TeV & 34/14 \\
&  & $I_0 \left(\frac{E}{E_0}\right)^{-\Gamma + \beta\log (E/E_0)}$ & $19.3\pm 0.3\pm 0.2$ & $2.31\pm 0.01\pm0.01$ & $\beta= 0.076 \pm 0.009\pm 0.008$ & 45/14 \\

\doublerule 
& & $I_0 \left(\frac{E}{E_0}\right)^{-\Gamma}$ & $15.0\pm 0.5\pm 2$ & $2.23\pm 0.02\pm 0.04$ & - & 39/16 \\
B & $0.8^{\circ}$ & $I_0 \left(\frac{E}{E_0}\right)^{-\Gamma}\exp\left(-\frac{E}{E_c}\right)$ & $16.1\pm 0.6\pm 2.$ & $2.06\pm 0.05\pm 0.08$ & $E_c = 15\pm 5\pm 6$ TeV & 18/15 \\
& & $I_0 \left(\frac{E}{E_0}\right)^{-\Gamma + \beta\log (E/E_0)}$ & $16.5\pm 0.6\pm 2.$ & $2.21\pm 0.03\pm 0.04$ & $\beta= 0.08 \pm 0.02\pm 0.03$ & 21/15 \\
\doublerule
& & $I_0 \left(\frac{E}{E_0}\right)^{-\Gamma}$ & $19.8\pm 0.4$ & $2.38\pm 0.02$ & - & 40.4/15 \\
H.E.S.S. 2006 & $0.8^{\circ}$ & $I_0 \left(\frac{E}{E_0}\right)^{-\Gamma}\exp\left(-\frac{E}{E_c}\right)$ & $21.0\pm 0.5$ & $2.26\pm 0.03$ & $E_c = 24.8\pm 7.2$ TeV & 16.9/14 \\
&  & $I_0 \left(\frac{E}{E_0}\right)^{-\Gamma + \beta\log (E/E_0)}$ & $21.0\pm 0.4$ & $2.29\pm 0.02$ & $\beta= -0.17 \pm 0.04$ & 14.5/14 \\

\hline
\end{tabular}
\caption{Fit parameters for various fits to the nebula spectrum extracted from a symmetric region of $0.8^{\circ}$ and $0.4^{\circ}$ radius, respectively, with $E_0 = 1$TeV and $I_0$ in units of $10^{-12}\mathrm{cm}^{-2}\mathrm{s}^{-1}\mathrm{TeV}^{-1}$. In all cases, the first errors quoted are statistical and the second errors are systematic. Curved models are preferred for the results from analysis A, fitted in the energy ranges [0.133,91] TeV in the core region, [0.2, 91] TeV in the $0.8^\circ$ radius, and for the shorter exposure analysis B in the energy range [0.14, 91] TeV. The fit results from \cite{Funk06} are also provided for comparison.}
\label{tab:spectralfits}
\end{center}
\end{table*}

\section{VHE $\gamma$-ray nebula HESS\,J1825--137}

Firstly, we present maps and spectra for comparison with previous H.E.S.S. results \citep{Funk06}.
Excess count maps of the region, constructed with analyses A and B using the ring background method, are shown in figure \ref{fig:sigmapAcorereg}, in which the binary system LS\,5039 is clearly visible as an additional point-like source \citep{deNaurois06}. The colour scale of figure \ref{fig:sigmapAcorereg} is optimised for HESS\,J1825--137, such that the image of LS\,5039 is saturated and appears larger than the PSF. The location of the more recently discovered hard-spectrum extended source HESS\,J1826--130 is indicated by a green dashed circle north of the nebula. The extended emission from this source overlaps with the larger region of spectral extraction, although the contribution to the flux is expected to be minor. The position of PSR\,B1823--13 is also shown. 
The point-spread function (PSF) was modelled using a triple-Gaussian function; for the map in figure \ref{fig:sigmapAcorereg}, the 68\% containment radius of the PSF is $0.064^{\circ}$, whilst for analysis B, it is slightly larger at $0.077^{\circ}$ as a result of the lower median energy. 
A correlation radius of $0.07^\circ$ was used for all excess count maps, and the significance was calculated using the number of ON and OFF events, as described in \cite{LiMa83}. 

A spectrum of the nebula, shown in figure \ref{fig:spectrum}, was extracted from analyses A and B using a spectral extraction region with a radius of $0.8^{\circ}$ that matches the spectrum used in \cite{Funk06}, and it was centred on the best-fit position of the nebula obtained from their 2D Gaussian morphological fit ($18^{h}25^{m}41^{s}$, $-13^\circ50'21''$). Below 1~TeV, the flux obtained with analysis B appears systematically lower than that of analysis A, although the two spectra are compatible within two standard deviations. This reduction in flux explains the comparatively harder spectral indices that are obtained with analysis B, and it may indicate difficulties with the background estimation at low energies. 

Additionally, a spectrum was extracted from the core region ( of $0.4^{\circ}$ radius)  of the nebula in order to characterise the spectrum of the brightest part of the nebula, which contains the peak and highest energy emission but decisively omits any contribution from HESS\,J1826--130. The results of the fits are given in table \ref{tab:spectralfits}. The fitted parameters agree reasonably well with those of \cite{Funk06}. 
The residuals of the spectral points to the fit were calculated as (point -- model) / model.

Curved spectral models are favoured over a power-law model for all three spectra shown in table \ref{tab:spectralfits} at the $7\sigma$ level for analysis A and at $3\sigma$ for analysis B. At all energies, the flux from the core region of HESS\,J1825--137 dominates that from HESS\,J1826--130 \citep{Ang1826ICRC17}. Towards the highest energies, the emission from the core region with $0.4^\circ$ radius of HESS\,J1825--137 converges towards the flux of the larger region of $0.8^\circ$ , which indicates that the region contains all of the high-energy emission. The integral flux above a given energy from HESS\,J1826--130 was found never to exceed 20\% of the flux from the $0.8^\circ$ region, such that we may assume that any contamination of HESS\,J1825--137 by HESS\,J1826--130 due to overlapping regions is not significant compared to the errors shown. Towards lower energies and especially below $\sim 1.5~$TeV, the spectrum of HESS\,J1826--130 is considerably contaminated by HESS\,J1825--137. The level of contamination will be addressed by a forthcoming publication on HESS\,J1826--130.

For the regions of $0.8^{\circ}$ radius, the $\chi^2 / \mathrm{ndf}$ indicates a marginal preference for a power law with an exponential cut-off over a log-parabola fit (\cite{Funk06}, whilst there is no preference shown for the core region of $0.4^{\circ}$ radius with analysis A. (Spectral points for all three spectra in table \ref{tab:spectralfits} are available as part of the supplementary information. )

\section{Morphological analysis}
\label{sec:morph}
\begin{figure*}
\centering
\makeatletter\ifaa@referee\relax\else
\begin{overpic}[width=\columnwidth]{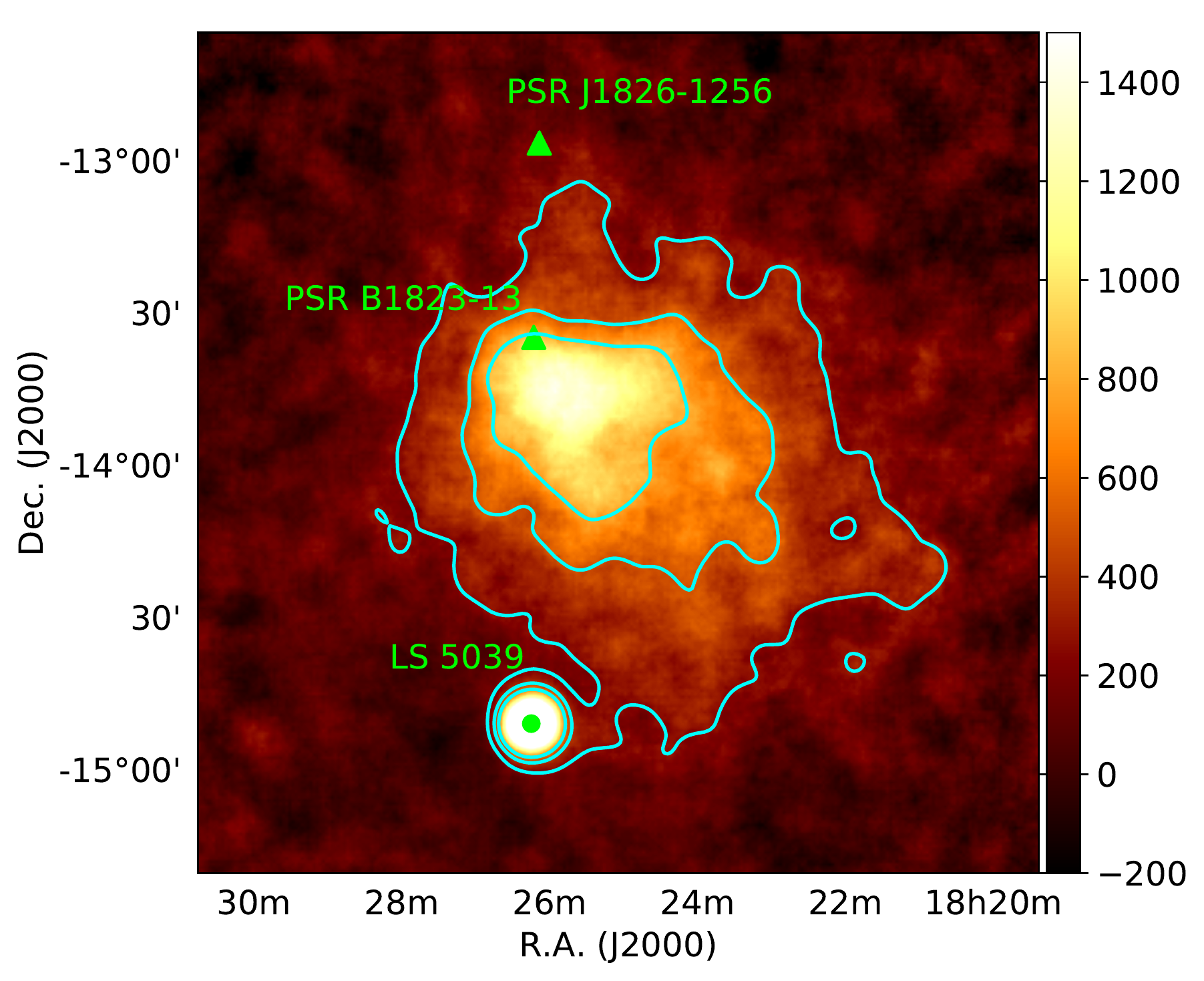}
\put(55,11){\Large{\textcolor{white}{E $< $ 1 TeV}}}
\end{overpic}
\begin{overpic}[width=\columnwidth]{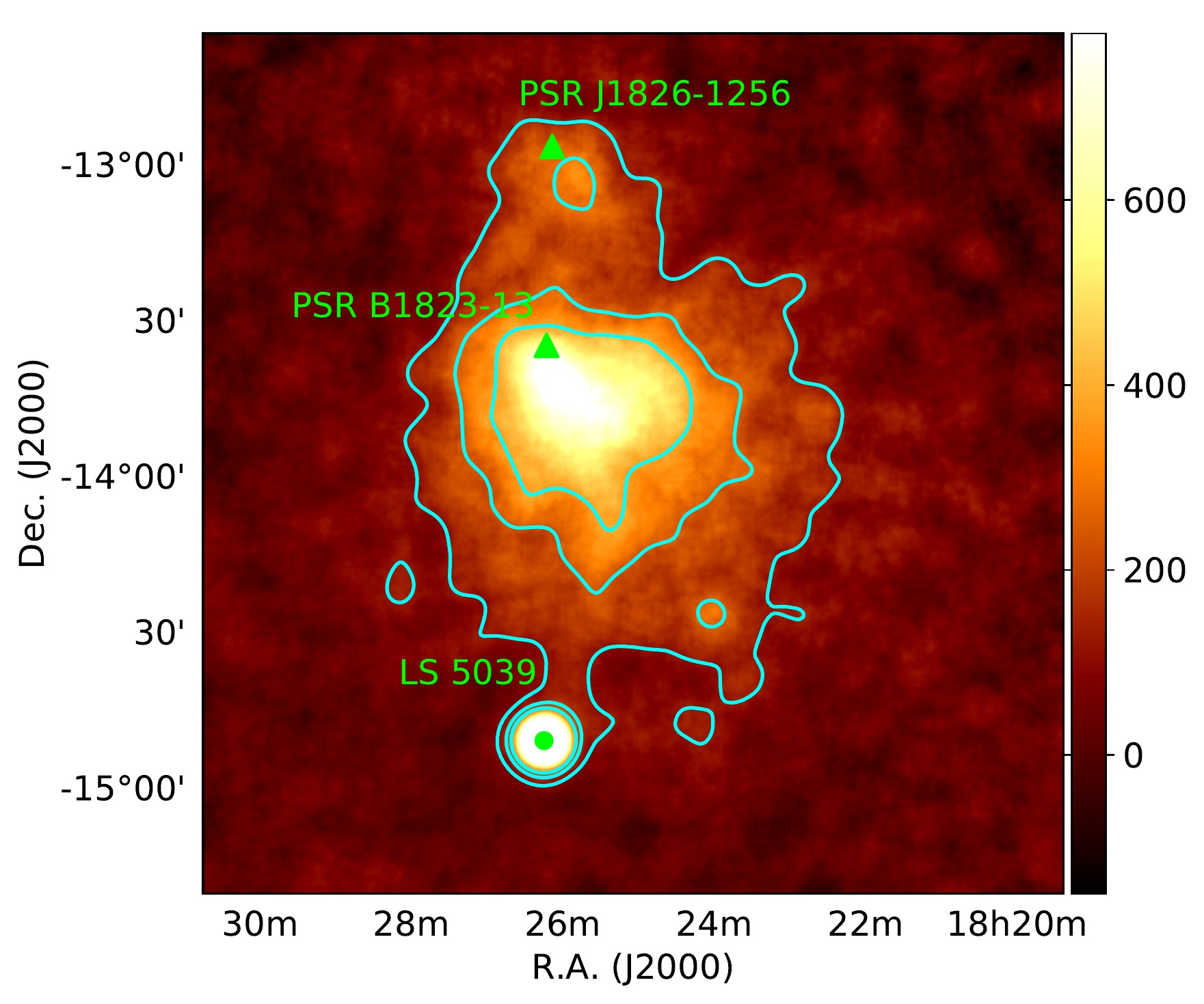}
\put(35,11){\Large{\textcolor{white}{1 TeV $<$ E $<$ 10 TeV}}}
\end{overpic}
\begin{overpic}[width=\columnwidth]{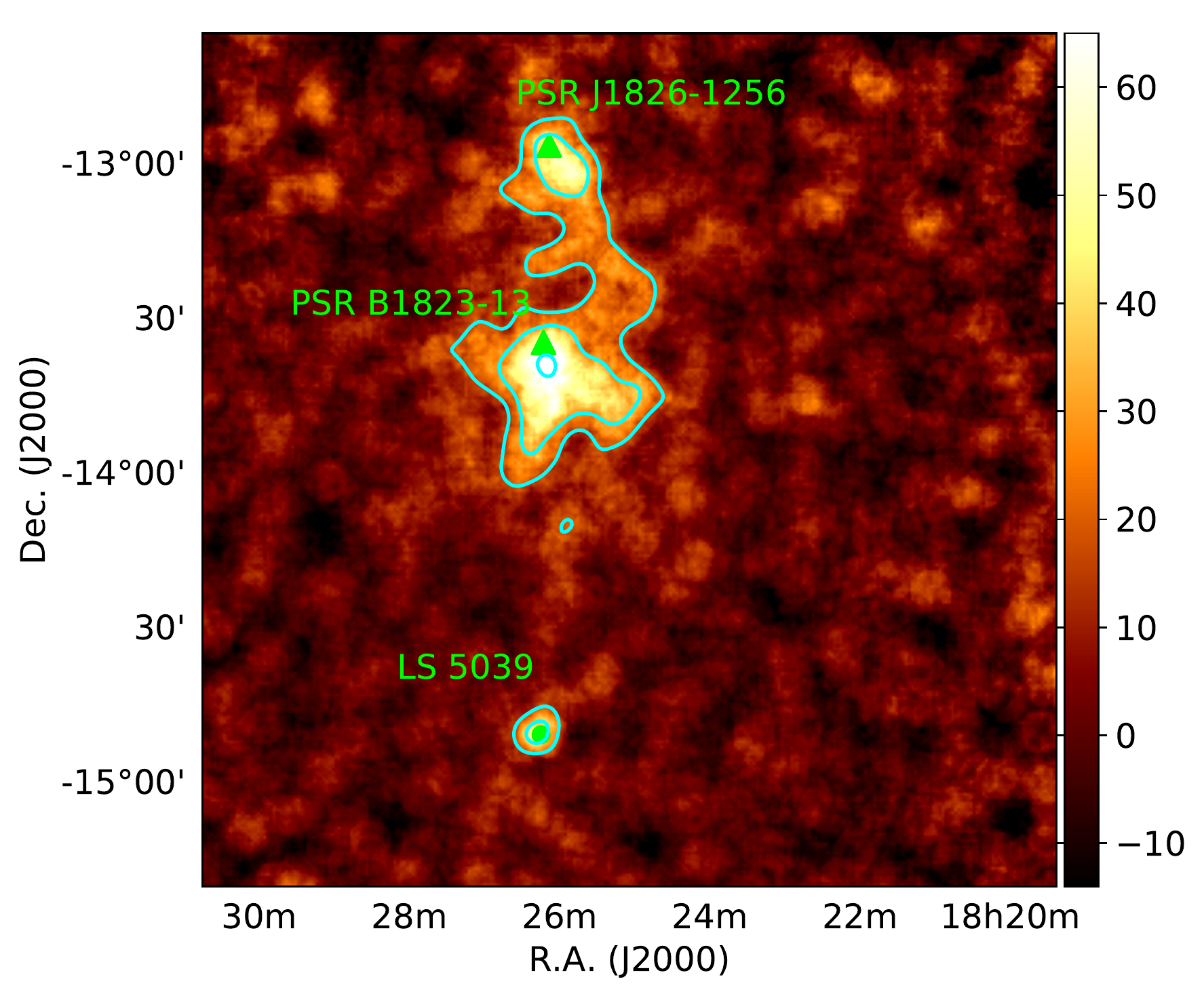}
\put(55,15){\Large{\textcolor{white}{E $>$ 10 TeV }}}
\end{overpic}
\begin{overpic}[width=\columnwidth]{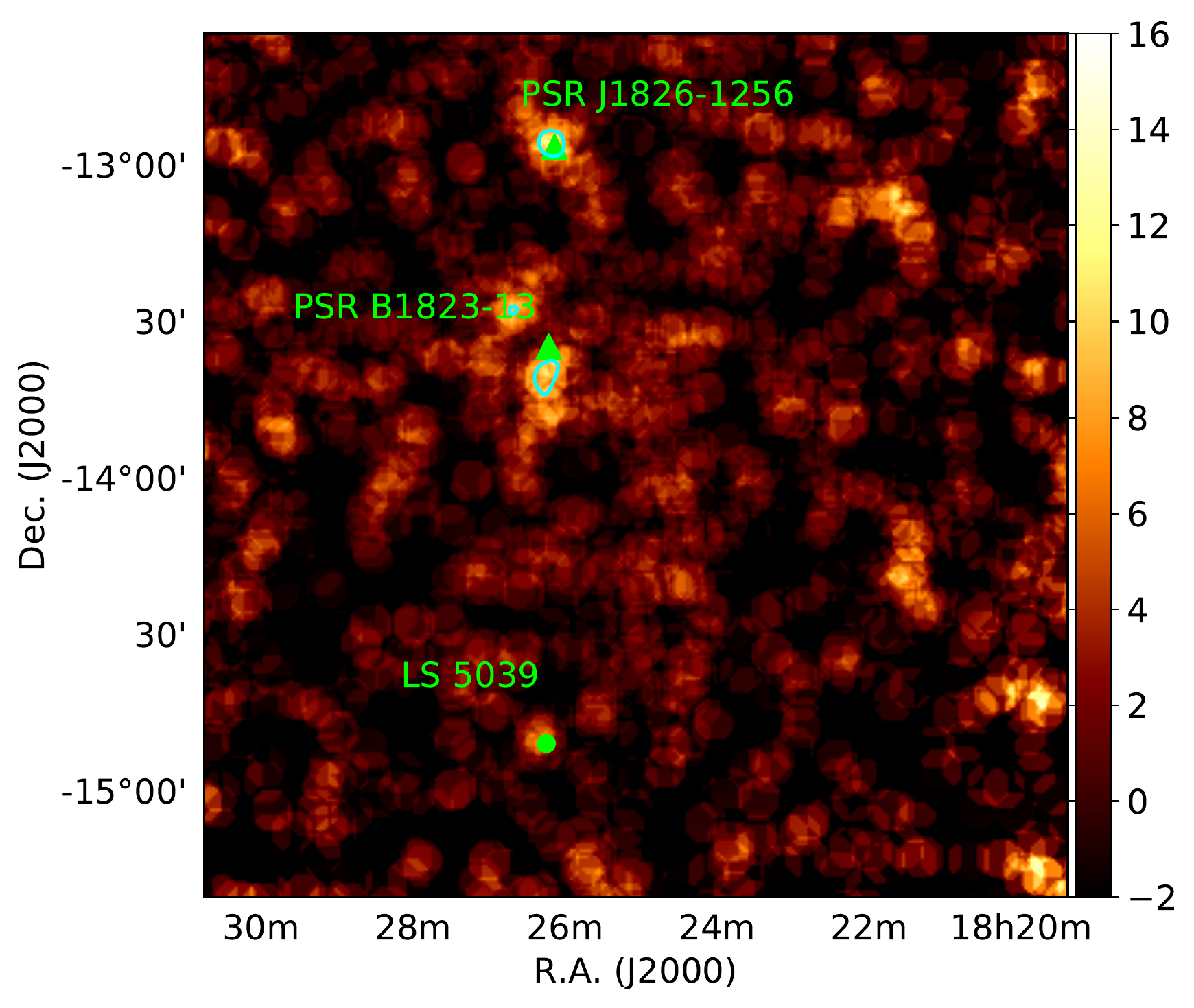}
\put(55,15){\Large{\textcolor{white}{E $>$ 32 TeV}}}
\end{overpic}
\fi\makeatother
\caption{Excess count maps of the HESS\,J1825--137 region in four different energy bands: E $<1$~TeV, $1$~TeV $<$ E $< 10$~TeV, E $> 10$~TeV, and E $> 32$~TeV. The size of the source is clearly much reduced at high energies. Other sources within the field of view include the binary LS~5039 and the hard-spectrum source HESS\,J1826--130. The positions of the pulsars PSR\,B1823--13 associated with HESS\,J1825--137 and PSR\,J1826--1256 (which might be associated with HESS\,J1826--130) are also shown. Significance contours are shown at 5, 10, and 15 $\sigma$ for maps with energies below 10~TeV, and at 3, 5, and 10 $\sigma$ for maps with energies above 10~TeV.}
\label{fig:Ebands}
\end{figure*}

Analysis A with the longer exposure (using CT1-4) was conducted in four energy bands, E $<1$~TeV, 1~TeV $<$ E $<$10~TeV, E $>10$~TeV, and E $> 32$~TeV, in which the nebula size can be clearly seen to decrease with increasing energy, as shown in figure \ref{fig:Ebands}. This is supporting evidence that the emission is attributable to PSR\,B1823--13, and it provides some indication that the electron population cools over time as the particles are transported away from the pulsar. The peak of the nebula emission is also seen to be offset from the pulsar position until energies above 32~TeV are reached.

Overall, the nebula morphology is highly asymmetric, extending a large distance away from the pulsar towards the south, but with only minor extension towards the north. 
The changing extent with energy is apparent in the excess count maps in different energy bands shown in figure \ref{fig:Ebands}. At the highest energies (E $>$ 32 TeV), the nebula is significantly less extended and reduced to a small, almost symmetric emission region close to the pulsar. 
The presence of HESS\,J1826--130 becomes apparent in the medium-energy range, and the identification as an independent source is verified based on the spatial separation between HESS\,J1826--130 and HESS\,J1825--137 towards the highest energies, where an association of HESS\,J1826--130 with PSR\,J1826--1256 seems plausible \citep{Ang1826ICRC17}. PSR\,J1826--1256 is located at R.A.: $18^{h}26^{m}08.5^{s}$, Dec: $-12^\circ56'33''$, with spin-down age $\tau = 1.44\times 10^{4}$~yr, a spin-down power of $\dot{E}=3.6\times 10^{36}~\mathrm{erg~s}^{-1}$ and a period of $P=0.110$~s, situated at a distance of 1.55~kpc \citep{Manchester05}.
Whilst there are some indications for emission from the binary system LS\,5039 at E $> 32$~TeV, this does not reach the $3\sigma$ level. It is notable that both HESS\,J1825--137 and HESS\,J1826--130 are, however, still present in the excess map above 32~TeV.

\begin{figure}
\centering
\includegraphics[width=\columnwidth]{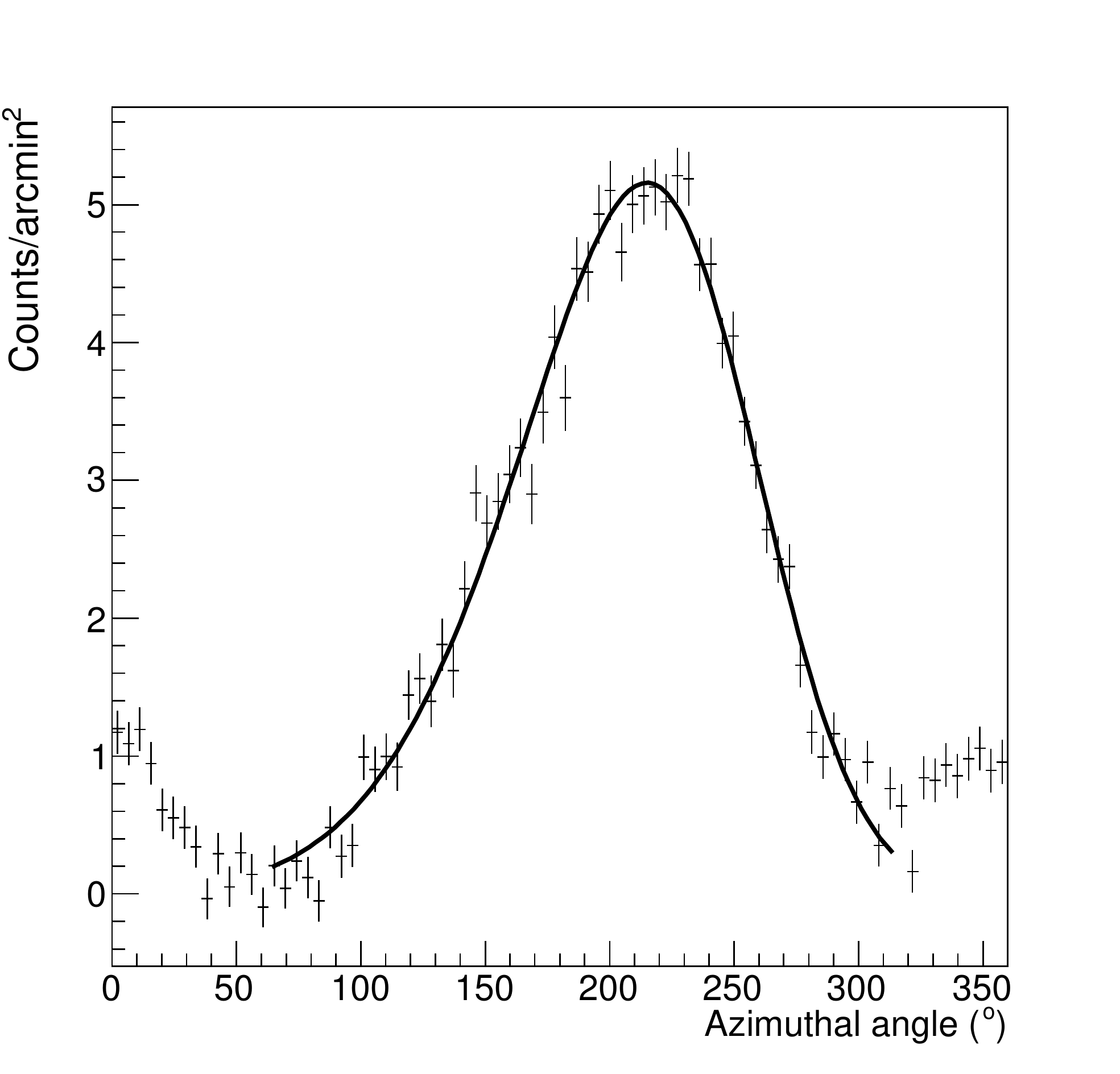}
\caption{Azimuthal profile of emission around the position of PSR\,B1823--13 out to a radial distance of $1.6^\circ$. The emission profile is well described by an asymmetric Gaussian between the two minima.}
\label{fig:aziprof}
\end{figure}

\subsection{Azimuthal profile}
\label{sec:aziprofile}

To explore the nebula morphology in more detail, the distribution of emission in the nebula was profiled as a function of azimuthal angle around the pulsar position. The azimuthal angle was measured anti-clockwise around the pulsar from north, and the profile included excess emission out to a radius of 1.6$^\circ$ from the pulsar. Masks were applied to exclude the emission from the other sources in the region. 
This distribution hints at the existence of a preferred axis for the particle flow into the nebula, with some spread on either side of this.
It was found that the azimuthal distribution of $\gamma$-ray counts per arcmin$^2$ is well described by an asymmetric Gaussian, as shown in figure \ref{fig:aziprof}, which was used to find the major axis of the emission. The range of the fit was restricted to fall between two minima of the distribution, which we found using a weighted moving-average approach to be located at $63^\circ$ and $315^\circ$ , respectively. 

The fitted mean was found to lie at a value of $208.0^\circ \pm 0.6^\circ_{\mathrm{stat.}} \pm 10.^\circ_{\mathrm{sys.}}$ using analysis A, where the systematic error arises from the differences between analysis pipelines. For analysis B, the mean of the asymmetric Gaussian fit was found at an azimuthal angle of $211^{\circ} \pm 1^\circ_{\mathrm{stat.}} \pm 6^\circ_{\mathrm{sys.}}$, well within the errors of the method. This angle was used to define the major axis of the emission that emanates south of the pulsar. The minor axis of the emission was subsequently defined to be perpendicular to this. 

The orientation angle of $17^{\circ}\pm 12^{\circ}_{\mathrm{stat.}}$ (measured anti-clockwise from north) found by \cite{Funk06} using a 2D Gaussian morphological fit agrees within the errors with the major axis found above, corresponding to an orientation angle of $28.0^{\circ} \pm 0.6^\circ_{\mathrm{stat.}} \pm 10.0^\circ_{\mathrm{sys.}}$, taking into account the $180^\circ$ change of reference. 

\begin{figure*}
\centering
\includegraphics[width=0.68\columnwidth]{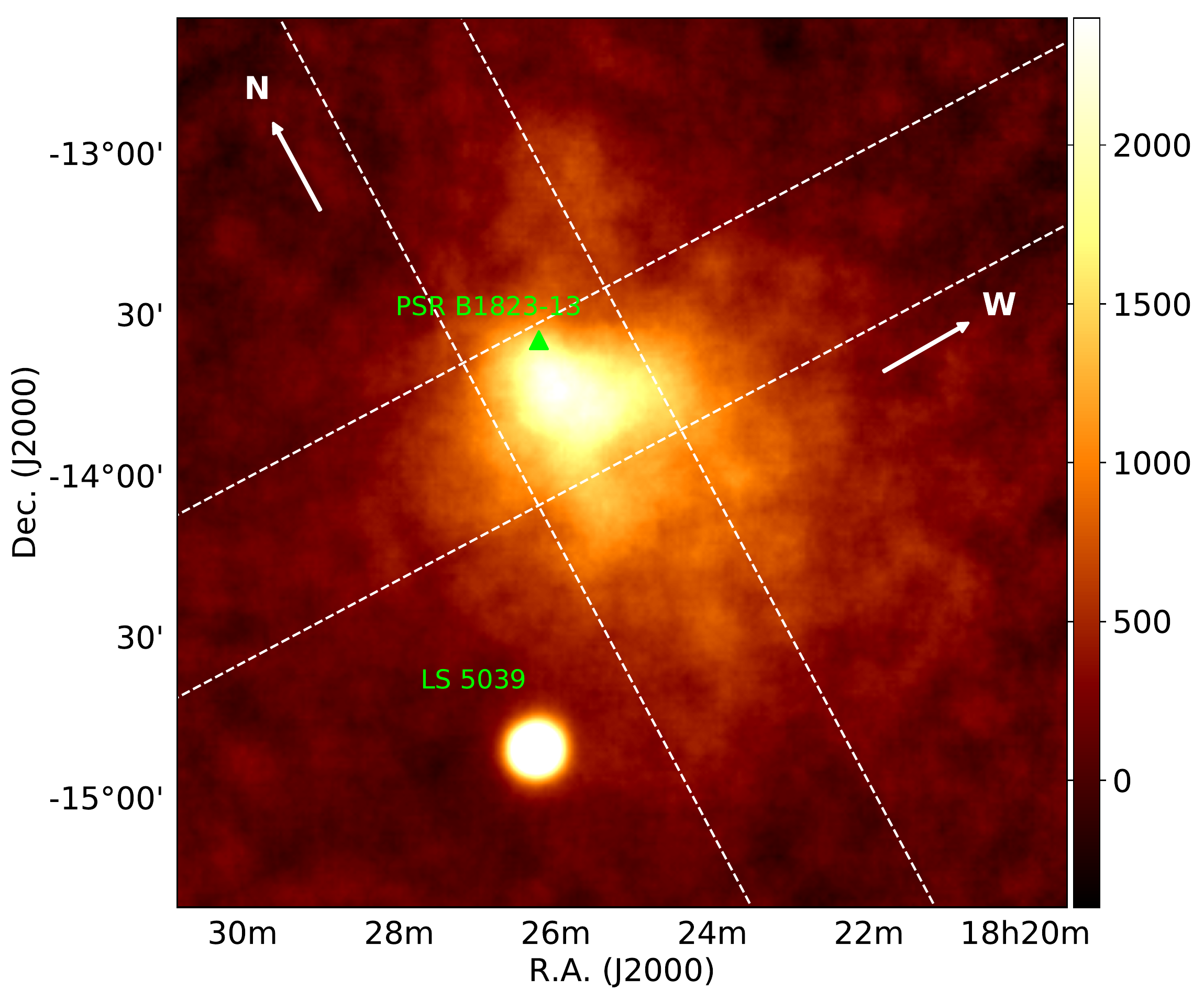}
\includegraphics[width=0.65\columnwidth]{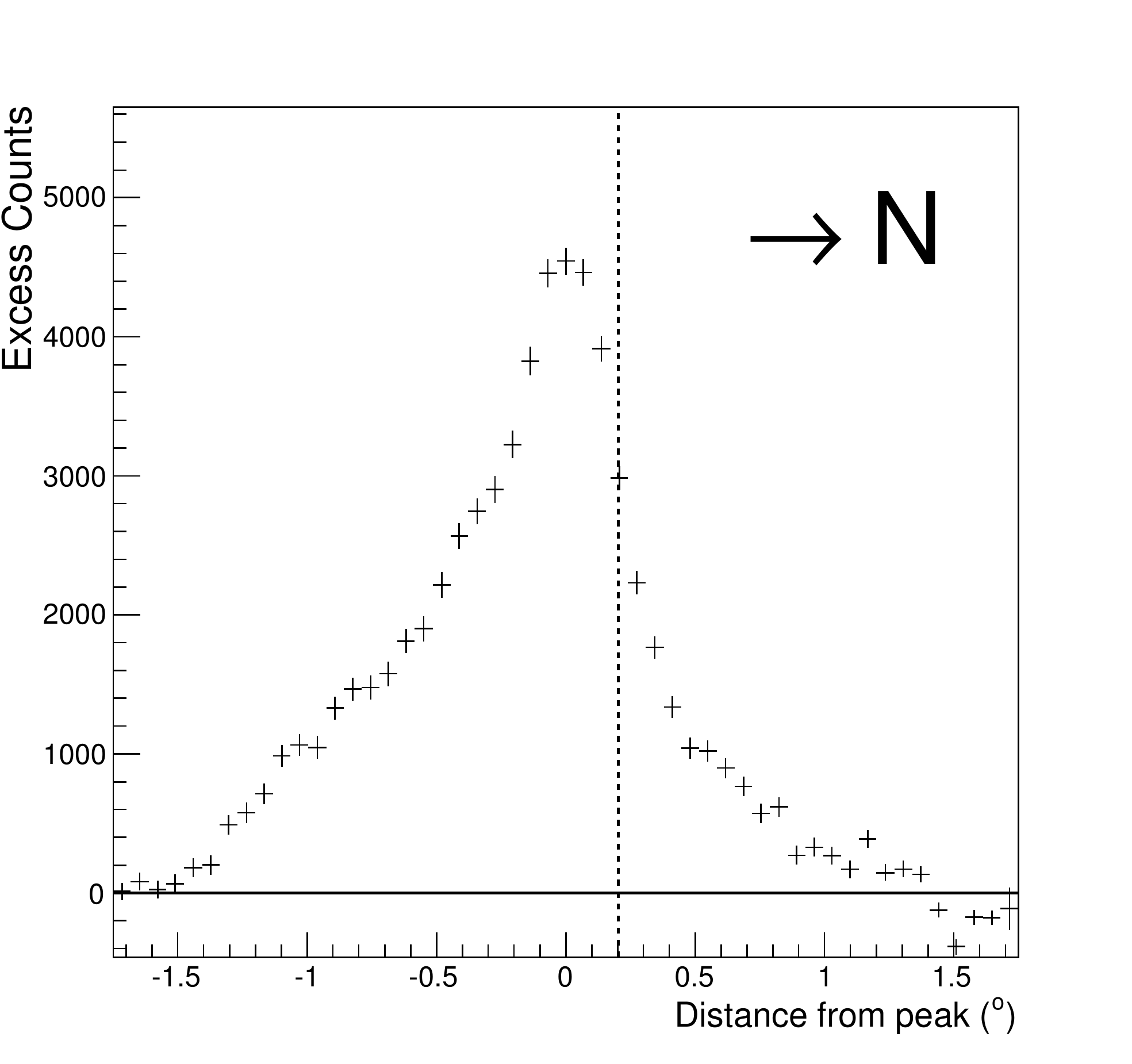}
\includegraphics[width=0.65\columnwidth]{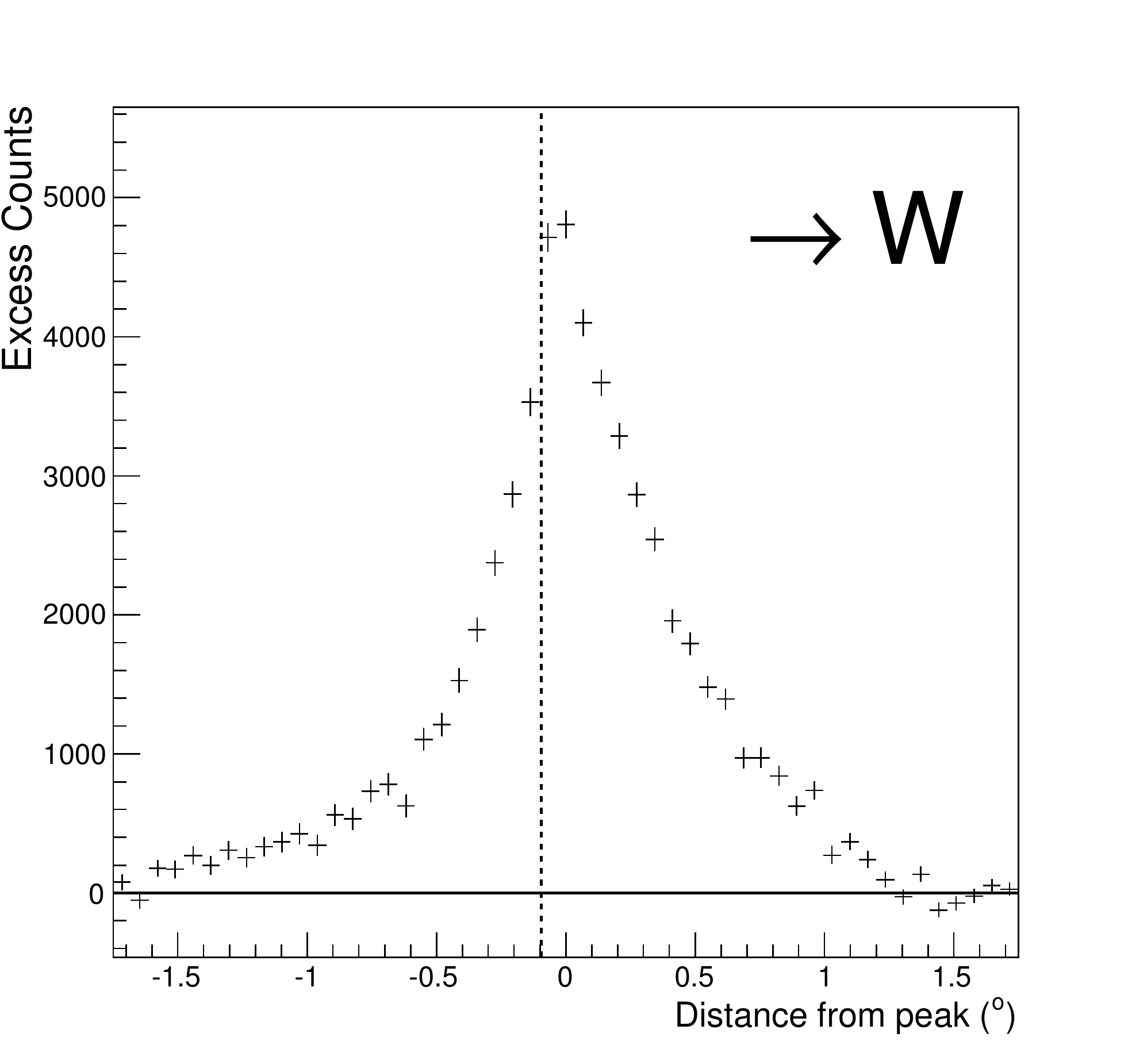}

\caption[Slices along the major and minor axes from the excess map]{Slices taken along the major and minor axes from the uncorrelated excess map using analysis A, centred on the peak position, with dimensions $3.5^\circ \times 0.5^\circ$. The location of the pulsar along the slice is indicated by a dashed line, and the emission peak is clearly offset towards the south, with the emission extending out to $\sim1.5^\circ$ from the pulsar. In contrast, the emission drops off extremely steeply from the emission peak north of the pulsar. South and west are orientated towards lower right ascension values. }
\label{fig:dataslices}
\end{figure*}

\subsection{Slice profiles}
\label{sec:slices}

Slices were taken from the excess map along the major and minor axes of the emission ($208^\circ$ and $118^\circ$) as determined from the azimuthal profiles, with dimensions $3.5^\circ \times 0.5^\circ$. These slices are shown for analysis A in figure \ref{fig:dataslices}; the position of the pulsar is indicated by a dashed line. The orientation of the slices is serendipitously away from either of the neighbouring sources such that no masks were applied to these slices. Emission from the $\gamma$-ray source HESS\,J1826--130 may contribute, but it is clearly a minor contribution along the slice width and does not lead to any visible features in either slice.

In the north-south direction, the steepness of the drop in emission towards the north is clearly apparent, whilst the emission towards the south seems to extend farther in analysis A than was previously seen in \cite{Funk06}. 
Towards the south, the emission profile becomes less steep from around $-0.5^\circ$ to $-1^\circ$ offset from the pulsar. This change in slope could be an indication of multiple components to the emission.

Gaussian fits to the excess slices were found to provide a poor fit to the total excess distribution along the slice, although this was improved slightly by an asymmetric Gaussian fit. This is particularly due to the increased steepness of the distribution from the peak of the emission towards the pulsar position. 
The rms of the distribution along the excess slices were $0.554^\circ \pm 0.005_{\mathrm{stat.}}^\circ \pm 0.05_{\mathrm{sys.}}^\circ$ and $0.512^\circ \pm0.005_{\mathrm{stat.}}^\circ \pm 0.05_{\mathrm{sys.}}^\circ$ along the north-south and east-west directions, respectively, a factor of 1.2 -- 1.3 larger than the Gaussian widths. (The Gaussian widths were $\sigma = 0.465^\circ \pm0.005_{\mathrm{stat.}}^\circ \pm 0.02_{\mathrm{sys.}}^\circ$ and $\sigma = 0.390^\circ \pm0.005_{\mathrm{stat.}}^\circ \pm 0.008_{\mathrm{sys.}}^\circ$ along the north-south and east-west directions, respectively, although with an offset from the peak position.) In general, these profiles are consistent with the picture of a pulsar wind that continuously injects particles into the nebula that are then transported preferentially towards the south. The particle population spreads out primarily along this direction; with spreading along the east-west direction (although with a non-Gaussian profile) rather more symmetric. 

\subsection{Determining the peak location}

The peak of the nebula emission is found to be at $18^{h}25^{m}49^{s} \pm 14_{\mathrm{stat}}^{s} \pm 5_{\mathrm{sys}}^{s}$, $-13^\circ46'35''\pm14_{\mathrm{stat}}''\pm1_{\mathrm{sys}}'$ by using the excess slices to determine the location of the peak emission of the nebula in an iterative process as follows. 
Both slices were always centred at the same position, initially at the pulsar location. 
The slice profiles were fit with a Gaussian function in a small range ($0.5^\circ$ total) around the maximum; the offset of the fitted mean from the centre of each slice was the shift to the peak. 
In a first step, we found a shift of $0.20^\circ \pm 0.02^\circ$ of the peak emission from the pulsar position along the north-south axis. 
After two iterations, there was no significant shift of the peak away from the centre of the slices along either axis. 
As shown in figure \ref{fig:dataslices}, the pulsar position is offset from the  final peak location by $0.20^\circ$ along the north-south axis and by $-0.09^\circ$ along the east-west axis.
For comparison, the best-fit position reported in \cite{Funk06} was $18^{h}25^{m}41^{s}\pm 3_{\mathrm{stat}}^{s}$, $-13^\circ 50'21''\pm 35_{\mathrm{stat}}''$; the two peak locations are compatible in Right Ascension and slightly offset in Declination, which may be a consequence of the different methods used to describe the morphology.

\begin{figure*}
\begin{center}
\makeatletter\ifaa@referee\relax\else
\begin{overpic}[trim=10mm 0mm 10mm 5mm,clip,width=1.1\columnwidth]{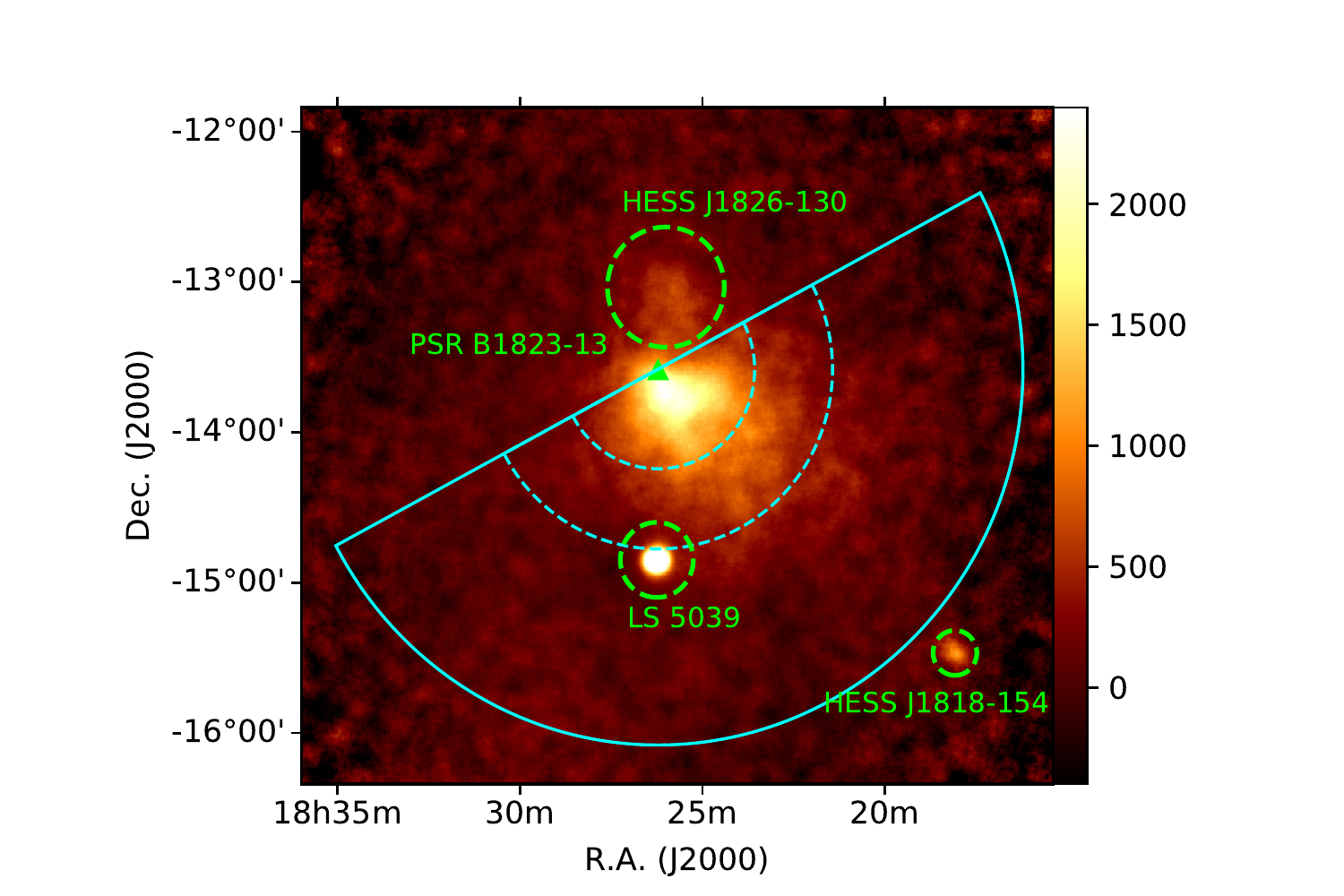}
\end{overpic}
\fi\makeatother
\includegraphics[width=0.9\columnwidth]{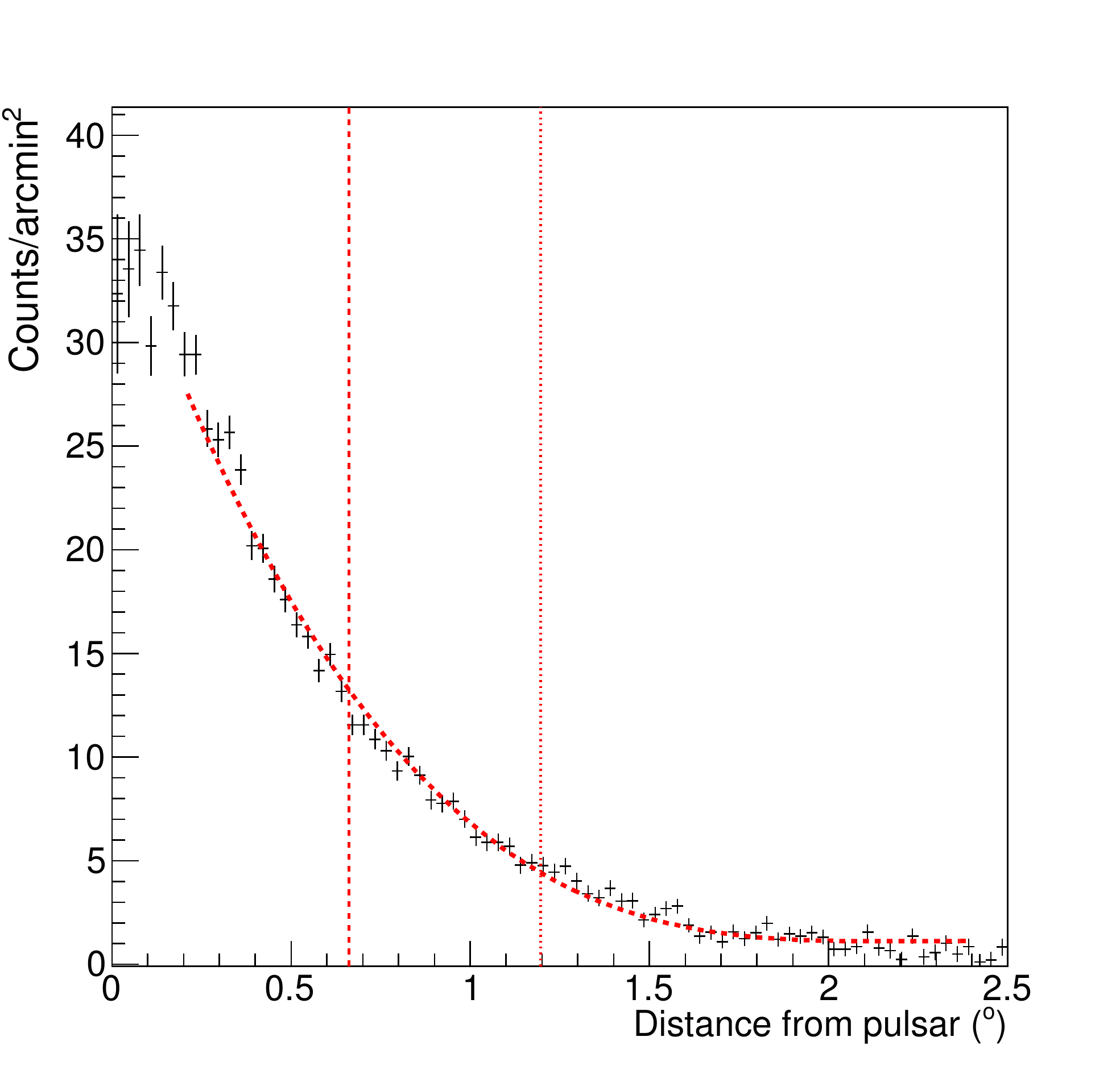}
\caption[Radial profile fitting for extent measurements]{Semi-circular region (left), with the pulsar at the apex and a mask applied over LS~5039, was extracted from the excess maps to form a radial profile of the emission (right), shown here for the full energy range with analysis A. A mask was also applied over HESS\,J1826--130, which is particularly relevant for measuring the northern extent. The radial profile was fitted by a polynomial (equation \eqref{eq:polfit}) to characterise the extent of the emission. Dashed and dotted lines mark the radial distance at which the emission drops to $1/e$ and $10~\%$ of the peak value, respectively.}
\label{fig:radialprofile}
\end{center}
\end{figure*}

\subsection{Nebula extent}
\label{sec:extentmethod}

The radial extent of the nebula was measured using the radial profile of the emission in the southern half of the nebula (south of the minor axis along $118^\circ$ as defined in section \ref{sec:aziprofile}), adopting an approach similar to that used in \cite{Eger16}. A mask of $0.25^{\circ}$ radius was applied over LS~5039 (HESS\,J1826--148) to avoid contamination of the profile of the excess nebula emission, and the radial profile was taken from the current pulsar position.
Rather than an immediate drop in emission away from the pulsar, the peak emission is roughly flat out to $\sim0.2^{\circ}$ radial distance, as shown in figure \ref{fig:radialprofile}. The extent of the emission was characterised by fitting a polynomial to the radial profile, in the range $0.2^{\circ} - 2.4^{\circ}$, according to

\begin{equation}
y = 
\begin{cases}
a(x-r_0)^n +c & ( x < r_0 ) \\
c & ( x \geq r_0 )
\end{cases}~,
\label{eq:polfit}
\end{equation}

\noindent such that with increasing $r$, the emission decreases out to a distance $r_0$ at which it approaches a constant value, $c$. Whilst the parameter $a$ simply provides the overall normalisation, to avoid a dependency on the order of the polynomial $n,$ the radius at which the fitted function dropped to a fixed fraction of the peak value ($1/e$, referred to as $r_{1/e}$) was used as a measure of the nebula extent and was found to be robust against the value of $n$, with $n=3$ chosen arbitrarily. A moving-average approach along the excess emission profile was used to find the radial offset and value of the peak of the emission (as in figure \ref{fig:radialprofile}), and the radial profile fitted from the peak out to large radii ($\sim2.4^\circ$). The peak of the emission was found to vary with energy between $0^\circ -0.2^\circ$ radius from the pulsar, shifting towards the pulsar at higher energies. 
The distance from the pulsar at which the fitted function evaluated to $1/e$ of the peak value was found to be not strongly dependent on the functional form used. Results obtained using an exponential function to describe the radial profile of the emission were consistent with those obtained using the polynomial equation \eqref{eq:polfit}.

The characteristic size of the nebula over the full energy range was found to be $r_{1/e} = 0.66^\circ \pm 0.03_{\mathrm{stat.}}^\circ \pm 0.04_{\mathrm{sys.}}^\circ$ from analysis A ($0.72^\circ \pm 0.03_{\mathrm{stat.}}^\circ \pm 0.04_{\mathrm{sys.}}^\circ$ from analysis B), which, when we adopt a distance of 4\,kpc to the nebula, corresponds to a physical extent of 46 pc. Over the full energy range, parameters $a$ and $c$ of \eqref{eq:polfit} were ($-3.5 \pm 0.2$) arcmin$^{-2}$deg$^{-n}$ and ($1.14 \pm 0.08$) arcmin$^{-2}$  for analysis A (($-1.0 \pm 0.1$) arcmin$^{-2}$deg$^{-n}$ and ($-0.13 \pm 0.06$) arcmin$^{-2}$  for analysis B).
On the northern side, when we adopt the same approach to characterise the extent, but with a mask of $0.4^\circ$ radius over HESS\,J1826--130, the characteristic size of the nebula is found to be significantly smaller, $r_{1/e} =0.41^\circ \pm 0.03_{\mathrm{stat.}}^\circ \pm 0.09_{\mathrm{sys.}}^\circ$ from analysis A ($0.42^\circ \pm 0.04_{\mathrm{stat.}}^\circ \pm 0.1_{\mathrm{sys.}}^\circ$ from analysis B), but still extended, as suggested both by the excess slices in figure \ref{fig:dataslices} and the significance contours in figure \ref{fig:Ebands}.

\section{Nebula extent - implications for particle transport}
\label{sec:extentransport}

Given the known energy-dependent morphology of the nebula, it is instructive to attempt to quantify this change in extent with energy more rigorously. To this end, we applied the extent-measuring approach outlined in section \ref{sec:extentmethod} to radial profiles of the nebula emission extracted from independent energy bands. The high brightness and long exposure on the nebula enabled us to split the data into nine energy bands, six of which were common to analyses A and B (see table \ref{tab:extents}). These bands were chosen such that each bound was twice the energy of the previous, with the highest and lowest bands for each analysis still containing statistically significant emission. The long exposure of analysis A enables a measurement at energies above 32~TeV to be included, whereas for the lower exposure analysis B that includes CT5, this was found not to be significant, and the highest energy band was kept at E $>$ 16~TeV. Nevertheless, the lower energy threshold of analysis B, provided by CT5, enables the lowest energy band to be split into two: 125 $<$ E $<$ 250~GeV and E $<$ 125~GeV. For analysis A the lowest energy band is set at E $<$ 250~GeV. 

Measurements of the nebula extent were made in each energy range as listed in table \ref{tab:extents}. We interpreted the extent-energy relation in terms of particle transport mechanisms within an energy range where transport mechanisms could plausibly dominate this dependence. 

If energy-dependent diffusion is the dominating transport mechanism and no cooling losses occur (which may be assumed for young PWNe), then the nebula would be expected to increase in size with energy if all particles were injected at a single instance in time as a result of the energy dependence of the diffusion coefficient. However, where cooling losses play a role, the nebula would become more compact towards higher energies. 
A maximum extent of the nebula may be expected to occur at the $\gamma$-ray energy that corresponds to the energy of the parent particle population at which the cooling time becomes equal to the age of the nebula \citep{AAV95}.

Advective transport mechanisms may also influence the transport. Several studies have shown that advective processes may dominate in the inner regions of pulsar wind nebulae, although both diffusion and advection are likely to contribute to the overall transport \citep{Porth16,TangChevalier12,Khangulyan17,VanEtten11}.

This variation of extent with energy, shown in figure \ref{fig:REmodellines}, may be described by a simple power-law relation within the energy range over which the nebula extent is expected to be dominated by particle cooling. This provides an insight into the particle transport processes within the nebula. 

\begin{table}
\begin{center}
\begin{tabular}{rp{2.8cm}p{2.8cm}}
Energy Range & Extent (A) & Extent (B) \\
\doublerule
$< 125$ GeV & -- & $0.37^\circ \pm 0.15^\circ \pm 0.3^\circ$ \\
$125 - 250$ GeV & -- & $0.63^\circ \pm 0.07^\circ \pm 0.07^\circ$ \\
$ < 250$ GeV & $0.66^\circ \pm 0.04^\circ \pm 0.3^\circ$ & -- \\
$250 - 500$ GeV & $0.76^\circ \pm 0.03^\circ \pm 0.2^\circ$ & $0.71^\circ \pm 0.09^\circ \pm 0.01^\circ$ \\
$500$ GeV $- 1$ TeV & $0.72^\circ \pm 0.02^\circ \pm 0.05^\circ$& $0.72^\circ \pm 0.05^\circ \pm 0.2^\circ$\\
$1 - 2$ TeV & $0.64^\circ \pm 0.02^\circ \pm 0.11^\circ$ & $0.62^\circ \pm 0.07^\circ \pm 0.4^\circ$ \\
$2 - 4$ TeV & $0.47^\circ \pm 0.04^\circ \pm 0.08^\circ$ & $0.51^\circ \pm 0.05^\circ \pm 0.1^\circ$ \\
$4 - 8$ TeV & $0.38^\circ \pm 0.04^\circ \pm 0.13^\circ$ & $0.33^\circ \pm 0.07^\circ \pm 0.04^\circ$ \\
$8 - 16$ TeV & $0.27^\circ \pm 0.07^\circ \pm 0.06^\circ$ & $0.30^\circ \pm 0.12^\circ \pm 0.3^\circ$ \\
$> 16$ TeV & -- & $0.22^\circ \pm 0.12^\circ \pm 0.2^\circ$\\
$16 - 32$ TeV & $0.19^\circ \pm 0.08^\circ \pm 0.14^\circ$ & -- \\
$> 32$ TeV & $0.14^\circ \pm 0.1^\circ \pm 0.05^\circ$ &  -- \\
\end{tabular}
\caption{Extent measurements as a function of energy for analyses A and B, with statistical and systematic errors. The extent is characterised by the radial distance from the pulsar at which the flux reduces to $1/e$ of the peak value in each energy band. }
\label{tab:extents}
\end{center}
\end{table}

\subsection{Determining the fit range}

The following discussion of the energy dependence of the size of a PWN assumes that the electron population is cooled, with $\gamma$-rays produced by IC scattering, where $E_\gamma \propto E_e^k$ can be used to relate the $\gamma$-ray and electron energies. Within the Thomson regime, $k=2$, whereas within the Klein-Nishina (KN) regime, $E_\gamma\approx E_e$ \citep{Blumenthal70}.
Simulations of a cooling, radiating, and diffusing electron population with the EDGE code \citep{Lopez17ICRC,Hahn15} were used to validate the analytic estimates applied here and in the following in two models of the Galactic radiation fields \citep{Porter17,Popescu17}. 

\begin{figure}
\begin{center}
\includegraphics[width=0.95\columnwidth]{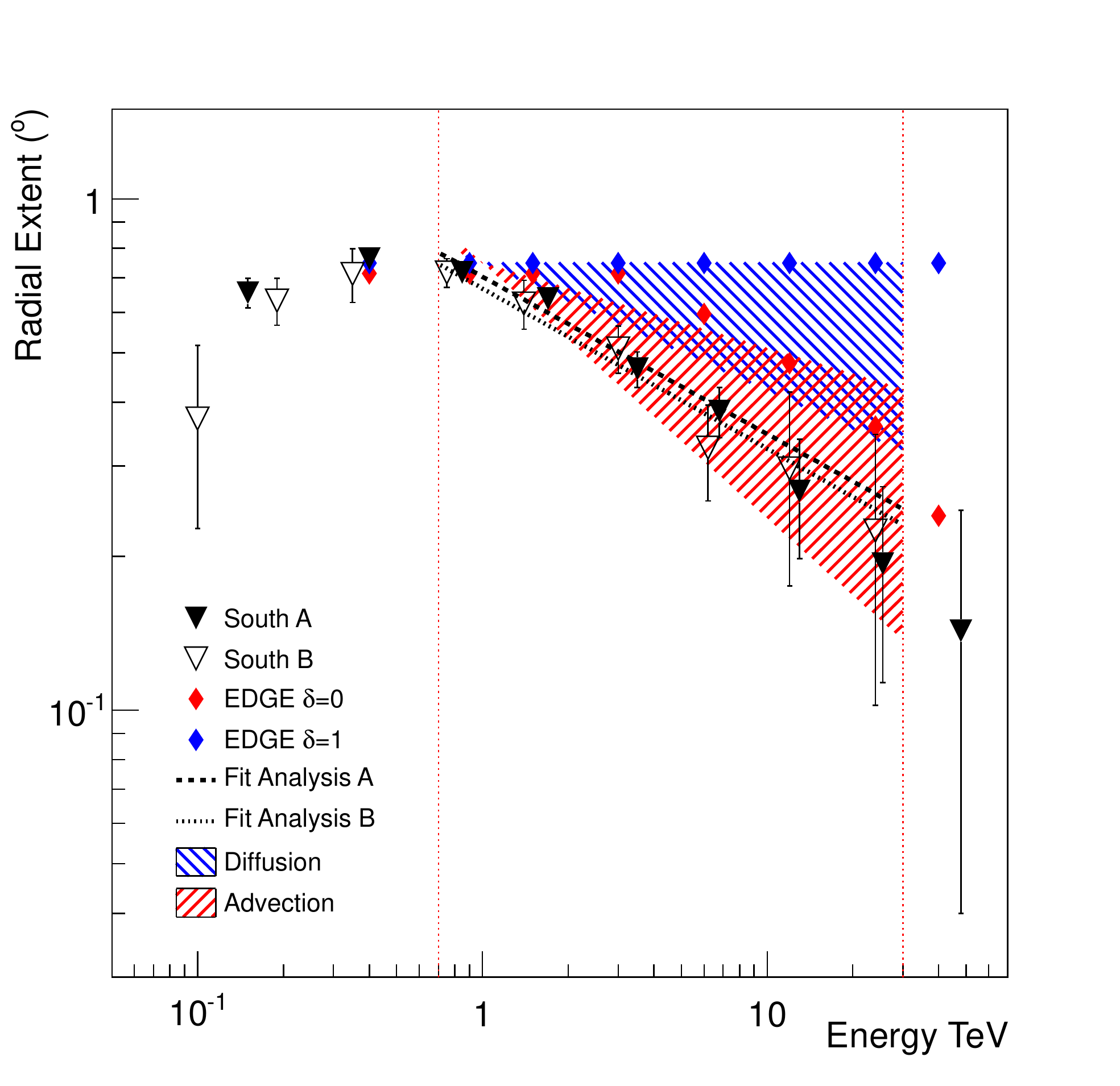}
\caption[Radius against energy gradients for different transport scenarios]{Variation of radial extent with energy. Shaded regions indicate compatible transport scenarios for IC scattering in the Thomson regime with $E_\gamma\propto E_e^2$, defined as $0 \leq \delta \leq 1$ for diffusion and $0 \leq \beta \leq 2$ for advection under the assumption of steady-state flow with constant density. The vertical dotted lines indicate the bounds of the fit at $\sim 700$~GeV and $\sim30$~TeV. A verification of the simple diffusion approximation using the EDGE code is also shown.} 
\label{fig:REmodellines}
\end{center}
\end{figure}

The radiation field model of \cite{Popescu17} for the Galactic location of HESS\,J1825-137 can be approximated by four black-body components;  the cosmic microwave background (CMB) with a temperature $T$ of 2.7\,K and an energy density $\omega$ of 0.25\,eVcm$^{-3}$; the far-infrared (FIR, dust, $T\,\sim\,40$\,K, $\omega\,\sim\,1\,\mathrm{eVcm}^{-3}$); near-infrared (NIR, T\,$\sim\,500$\,K, $\omega\,\sim\,0.4\,\mathrm{eVcm}^{-3}$), and visible light (VIS, $T\,\sim\,3500$\,K, $\omega\,\sim\,1.9\,\mathrm{eVcm}^{-3}$).
IC scattering in the TeV energy range was found to be dominated by the contribution from the FIR radiation field for both models.
EDGE was used to generate radial profiles for the nebula extent as a function of energy for the two bounding cases in the diffusion scenario. A magnetic field strength of $\sim5\,\mu$G was adopted for the simulated points shown in figure \ref{fig:REmodellines}, which include KN effects. No significant difference was found between the two Galactic radiation field models ( \cite{Porter17,Popescu17} in the resulting simulated nebula extent variation.

The lower bound of the fit corresponds to the apparent turn-over in figure \ref{fig:REmodellines} from around 0.7 TeV, which would correspond to a cooling break energy in the electron population of around 4\,TeV. This would require a magnetic field of $\sim12\,\mu$G, which is plausible for the earlier stages of nebula evolution, although X-ray measurements suggest a current magnetic field strength within the nebula closer to $\sim5\,\mu$G \citep{Uchiyama09}.

The upper bound on the fit range is given by the transition from FIR- to CMB-dominated IC scattering (see also figure \ref{fig:totalsed}). Using recent models of the interstellar radiation fields \citep{Porter17,Popescu17}, we found that the $E_\gamma \propto E_e^k$ relation does not remain constant over the fitted energy range. IC scattering of electrons off the FIR radiation field occurs in the transition region between the Thomson and KN regimes.

A power-law fit to the data was made, as shown in figure \ref{fig:REmodellines}, between these two bounds, and the relation $R=R_0(E_\gamma/E_0)^\alpha$ was adopted. The conclusions were unchanged, regardless of whether the highest energy point was included or omitted in the fit. The regions of this extent against energy plot that would be compatible with diffusion or advection are indicated in figure \ref{fig:REmodellines} with shaded areas, which are arbitrarily normalised to the lowest energy point within the fitted range. These two scenarios are discussed in the following sections, and the validity of our assumptions is discussed in section \ref{sec:discuss}.

\subsection{Diffusion}
\label{sec:diffusion}
Diffusive processes may be expected to dominate the particle transport once the pressure within the nebula has reduced to that of the surrounding ISM and the particles are no longer strictly confined within the nebula and begin to diffuse into the surrounding medium.
Under the assumption of diffusion and cooling losses, a power-law fit to figure \ref{fig:REmodellines} at energies above the maximum extent directly yields the diffusion index. Taking the radial extent $R$ to vary with the diffusion coefficient $D(E)$ and assuming that the cooling timescale $\tau$ is much less than the age of the nebula, we obtain
\begin{eqnarray}
R &=& \sqrt{2D(E)\tau} \\
 &=& \sqrt{2D_0\left(\frac{E_{e}}{E_{e0}}\right)^\delta \tau}~.
 \label{eq:diffradius}
\end{eqnarray}
When we assume that cooling losses for synchrotron and IC scattering vary with energy as $\tau \propto 1/E_{e}$, this yields a dependency of the nebula radius on the electron energy as $R \propto E_{e}^{(\delta - 1)/2}$. Given that the energy of $\gamma$-ray photons, $E_\gamma$, produced via IC-scattering interactions, varies with the electron energy, $E_{e}$, as $E_\gamma \propto E_{e}^2$ in the Thomson regime, the relation between $R$ and $E_\gamma$ applicable to figure \ref{fig:REmodellines} is

\begin{equation}
R = R_0 \left(\frac{E_\gamma}{E_{\gamma0}} \right)^{(\delta - 1)/4}~,
\label{eq:diffusionRE}
\end{equation}

\noindent where $R_0$ is the normalisation radius at $E_{0} = 1~$TeV.
Hence from the fitted index $\alpha$ of the slope in figure \ref{fig:REmodellines} (where $R \propto E^{\alpha}$), the energy dependence of the diffusion coefficient is obtained directly as $\delta = 4\alpha +1$. The diffusion index $\delta$ is expected to lie in the range $0.3-0.6$, with extremes of energy-independent diffusion at $\delta = 0$, and of Bohm diffusion at $\delta = 1$. The gradient of the fit in figure \ref{fig:REmodellines} directly yields the diffusion index, whilst the constant relates to the radius $R_0$. 

The fitted and derived parameters are given in table \ref{tab:fitextent}. The values of $\alpha$ from $-0.25$ to $0$ and from $-0.5$ to $0$ are compatible with diffusion in the Thomson and KN cases, respectively.
Negative values of the diffusion index $\delta$, as obtained for the Thomson case, are incompatible with the variation in spatial extent, and the photon energy is due to a diffusion-dominated particle transport scenario. 

\begin{table}
\begin{center}
\begin{tabular}{ccc}
Parameter & Value (A) & Value (B)\\
\doublerule 
$\alpha$ & $-0.29 \pm 0.04 \pm 0.05$ & $-0.29 \pm 0.06 \pm 0.1$\\
$R_0$ ($^\circ$) & $0.70 \pm 0.02 \pm 0.08$ & $0.69 \pm 0.04 \pm 0.2$ \\
$\delta$ (T) & $-0.16 \pm 0.15 \pm 0.2$ & $-0.17 \pm 0.24 \pm 0.1$\\
$\delta$ (KN) & $0.39 \pm 0.06 \pm 0.2$ & $0.4 \pm 0.1 \pm 0.5$\\
$\beta$ (T) & $0.7 \pm 0.2 \pm 0.3$ & $0.7 \pm 0.3 \pm 0.1$\\
$\beta$ (KN) & $2.3 \pm 0.3 \pm 0.8$ & $2.3 \pm 0.6 \pm 1$ 
\end{tabular}
\caption{Fitted parameters of the power-law relation, $R=R_0(E_\gamma/E_0)^\alpha$, fit to figure \ref{fig:REmodellines} and derived parameters for the diffusion and advection scenarios with analyses A and B for the Thomson (T) and Klein-Nishina (KN) dominated regimes. In all cases the first error provided is statistical and the second systematic. }
\label{tab:fitextent}
\end{center}
\end{table}

\subsection{Advection}
\label{sec:advection}

Bulk particle flow may, however, dominate the transport if the particle pressure within the nebula remains greater than the surrounding ISM pressure out to large distances, still with significant confinement. If advection is adopted as the dominant particle transport mechanism instead of diffusion, then a relation between the nebula radius $R$ and $\gamma$-ray energy $E_\gamma$ analogous to equation \eqref{eq:diffusionRE} can be obtained. During the particle outflow through the nebula, it is required that mass continuity is satisfied and that the flow follows a steady-state density profile $\dot{\rho} = 0$, such that the flow must preserve

\begin{equation}
\rho(r)A(r)v(r) = \mathrm{const.}~,
\label{eq:conserve}
\end{equation}

\noindent where $A(r)$ is the area through which the particles flow, and the radial dependence of $\rho(r)$, $A(r),$ and $v(r)$ is unknown. Assuming that the flow density $\rho$ is independent of the radius, then $v(r) \propto A(r)^{-1}$. Therefore, in the case of spherical symmetry, the area through which the flow travels is $A(r) \propto r^2$, and the flow velocity $v(r)\propto r^{-2}$. 
As the flow velocity is expected to vary with radius $r$ due to pressure on the nebula from the ambient medium, this can be parameterised as

\begin{equation}
v = v_0 \left(\frac{r}{r_0}\right)^{-\beta}~,
\label{eq:veladv}
\end{equation}

\noindent where $r_0$ and $v_0$ are the initial radius and velocity of the nebula, respectively, and $\beta$ describes the radial dependency required in order to yield a constant density profile. In extreme cases, $\beta$ can take on values of 0 (for constant velocity expansion) or 2 in the case of constant density. 
By separation of variables and integrating over all radii and up to the cooling time, we obtain the proportionality relation $R \propto E_e^{-\frac{1}{(1+\beta)}}$. Accounting for the dependance of the $\gamma$-ray energy on the electron energy, $E_\gamma \propto E_{e}^2$ for the Thomson regime, this relation becomes

\begin{equation}
R = R_0  \left(\frac{E_\gamma}{E_{\gamma0}}\right)^{-\frac{1}{2(1+\beta)}}~,
\label{eq:advectionRE}
\end{equation}

\noindent where $R_0$ is the normalisation radius at $E_{0} = 1~$TeV, such that the gradient of the power-law fit to figure \ref{fig:REmodellines} yields a measure of the radial dependency of $\beta = -\frac{1}{2\alpha} -1$. The derived $\beta$ obtained for the Thomson and KN cases is given in table \ref{tab:fitextent}.
Both cases indicate a significant dependence beyond the constant velocity case, with values of $\alpha$ from $-0.5$ to $-1/6$ and from $-1$ to $-1/3$ compatible with advection in the Thomson and KN cases, respectively. Under the Thomson assumption, a significant dependence beyond the constant velocity case is assumed, whilst for the pure KN case, a relation closer to a spherically expanding flow is found.

\begin{table}
\centering
\begin{tabular}{ccccc}
Box & d(pc) & $\Gamma$ & $\Phi (1-5\,\mathrm{TeV})$ & $\sigma$\\
\doublerule
1 & 51 & 2.32 $\pm$ 0.05 $\pm$ 0.03 & 4.3 $\pm$ 0.2 $\pm$ 0.2 & 28.8 \\ %
2 & 41 & 2.21 $\pm$ 0.04 $\pm$ 0.03 & 5.9 $\pm$ 0.2 $\pm$ 0.3 & 34.2\\ %
3 & 36 & 2.24 $\pm$ 0.06 $\pm$ 0.03 & 4.3 $\pm$ 0.2 $\pm$ 0.2 & 24.6\\ %
4 & 41 & 2.29 $\pm$ 0.04 $\pm$ 0.03 & 6.0 $\pm$ 0.2 $\pm$ 0.2 & 38.8\\ %
5 & 26 & 2.13 $\pm$ 0.03 $\pm$ 0.03 & 9.6 $\pm$ 0.3 $\pm$ 0.4 & 51.4\\ %
6 & 18 & 2.09 $\pm$ 0.04 $\pm$ 0.02 & 6.9 $\pm$ 0.3 $\pm$ 0.3 & 35.05\\ %
7 & 36 & 2.20 $\pm$ 0.04 $\pm$ 0.02 & 5.4 $\pm$ 0.2 $\pm$ 0.2 & 33.7\\ %
8 & 18 & 2.06 $\pm$ 0.03 $\pm$ 0.01 & 10.4 $\pm$ 0.3 $\pm$ 0.3 & 53.6 \\ %
9 & 0 & 2.00 $\pm$ 0.03 $\pm$ 0.02 & 8.7 $\pm$ 0.3 $\pm$ 0.3 & 41.6\\ %
10 & 41 & 2.29 $\pm$ 0.07 $\pm$ 0.05 & 2.9 $\pm$ 0.2 $\pm$ 0.2 & 20.7\\ %
11 & 26 & 2.20 $\pm$ 0.05 $\pm$ 0.03 & 4.3 $\pm$ 0.2 $\pm$ 0.1 & 26.4\\ %
12 & 18 & 2.06 $\pm$ 0.06 $\pm$ 0.04 & 4.0 $\pm$ 0.2 $\pm$ 0.2 & 20.5\\ %
a & 106 & 3.3 $\pm$ 0.2 $\pm$ 0.2 & 0.5 $\pm$ 0.1 $\pm$ 0.01 & 8.8\\ 
b & 98 & 3.2 $\pm$ 0.2 $\pm$ 0.8 & 0.6 $\pm$ 0.2 $\pm$ 0.4 & 6.9 \\ 
c & 93 & 3.3 $\pm$ 0.3 $\pm$ 0.3 & 0.5 $\pm$ 0.2 $\pm$ 0.3 & 6.9 \\ 
d & 103 & 3.1 $\pm$ 0.3 $\pm$ 0.4 & 0.5 $\pm$ 0.2 $\pm$ 0.01 & 6.2\\ 
e & 91 & 2.90 $\pm$ 0.13 $\pm$ 0.05 & 1.2 $\pm$ 0.2 $\pm$ 0.1 & 12.8 \\ 
f & 81 & 2.8 $\pm$ 0.1 $\pm$ 0.1 & 1.3 $\pm$ 0.2 $\pm$ 0.2 & 13.3 \\ 
g & 75 & 2.7 $\pm$ 0.1 $\pm$ 0.1 & 1.3 $\pm$ 0.2 $\pm$ 0.2 & 11.1 \\ 
h & 73 & 2.9 $\pm$ 0.2 $\pm$ 0.1 & 0.8 $\pm$ 0.2 $\pm$ 0.1 & 8.2\\ 
i & 91 & 3.1 $\pm$ 0.2 $\pm$ 0.1 & 0.7 $\pm$ 0.2 $\pm$ 0.2 & 9.4\\ 
j & 77 & 2.61 $\pm$ 0.09 $\pm$ 0.07 & 2.1 $\pm$ 0.2 $\pm$ 0.2 & 17.7\\ 
k & 65 & 2.43 $\pm$ 0.07 $\pm$ 0.04 & 2.8 $\pm$ 0.2 $\pm$ 0.1 & 20.4\\ 
l & 57 & 2.40 $\pm$ 0.07 $\pm$ 0.03 & 3.1 $\pm$ 0.2 $\pm$ 0.2 & 21.0\\ 
m & 56 & 2.40 $\pm$ 0.10 $\pm$ 0.05 & 2.2 $\pm$ 0.2 $\pm$ 0.1 & 14.5\\ 
n & 57 & 2.45 $\pm$ 0.15 $\pm$ 0.3 & 1.3 $\pm$ 0.2 $\pm$ 0.1 & 8.7\\ 
o & 98 & 2.8 $\pm$ 0.2 $\pm$ 0.1 & 0.9 $\pm$ 0.2 $\pm$ 0.1 & 9.7 \\ 
p & 81 & 2.72 $\pm$ 0.11 $\pm$ 0.09 & 1.6 $\pm$ 0.2 $\pm$ 0.2 & 15.1\\ 
q & 65 & 2.33 $\pm$ 0.07 $\pm$ 0.02 & 3.0 $\pm$ 0.2 $\pm$ 0.2 & 21.2\\ 
r & 41 & 2.59 $\pm$ 0.1 $\pm$ 0.03 & 1.9 $\pm$ 0.2 $\pm$ 0.2 & 13.8 \\ 
s & 75 & 2.8 $\pm$ 0.1 $\pm$ 0.1 & 1.1 $\pm$ 0.2 $\pm$ 0.2 & 13.7\\ 
t & 57 & 2.3 $\pm$ 0.07 $\pm$ 0.02 & 3.2 $\pm$ 0.2$\pm$ 0.2 & 22.4 \\ 
u & 56 & 2.28 $\pm$ 0.08 $\pm$ 0.07 & 2.6 $\pm$ 0.2 $\pm$ 0.3 & 19.2\\ 
v & 57 & 2.47 $\pm$ 0.13 $\pm$ 0.06 & 1.4 $\pm$ 0.2 $\pm$ 0.1 & 12.4\\ 
w & 65 & 2.45 $\pm$ 0.16 $\pm$ 0.1 & 1.3 $\pm$ 0.2 $\pm$ 0.2 & 10.7 \\ 
x & 51 & 2.71 $\pm$ 0.16 $\pm$ 0.2 & 1.0 $\pm$ 0.2 $\pm$ 0.2 & 10.6 \\ 
y & 41 & 2.5 $\pm$ 0.2 $\pm$ 0.1 & 1.2 $\pm$ 0.2 $\pm$ 0.1 & 9.8 \\ 
z & 36 & 2.6 $\pm$ 0.2 $\pm$ 0.09 & 1.0 $\pm$ 0.2 $\pm$ 0.1 & 8.1\\ 
\end{tabular}
\caption{Spectral parameters from a power-law fit to each of the 38 boxes independently. The integrated flux from $1-5$~TeV is given in units of  $10^{-13}\mathrm{cm}^{-2}\mathrm{s}^{-1}$. For all results, the first error provided is statistical and the second systematic. The projected distance of the centre of each box to the pulsar is given for an assumed distance of 4\,kpc; the significance of the spectrum obtained from each box is also provided. }
\label{tab:spectralboxes}
\end{table}

\begin{figure*}
\centering
\includegraphics[width=1.75\columnwidth]{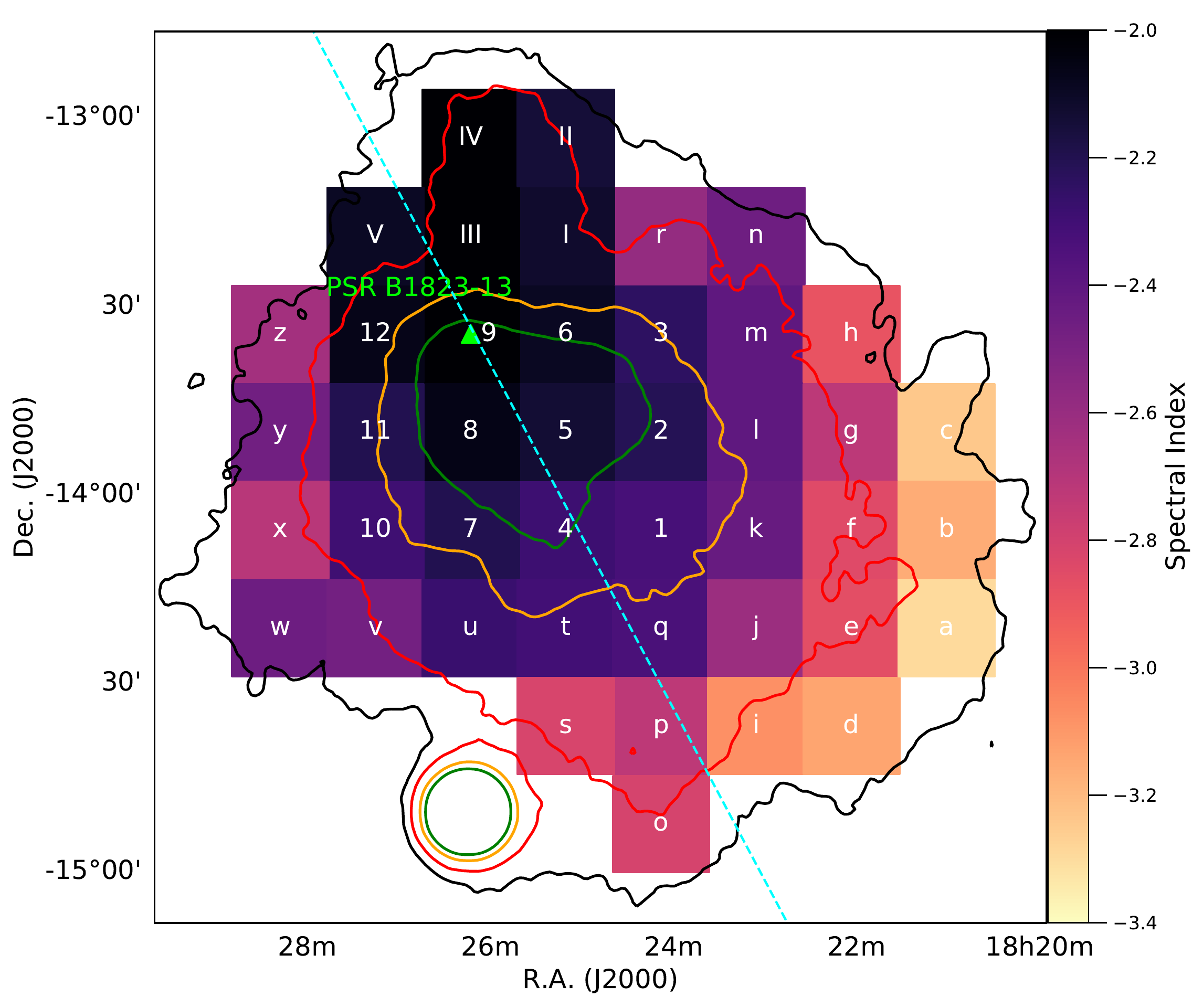}
\caption{Map of HESS\,J1825--137 showing the location and fitted power-law index of each of the 38 boxes used in independent spectral analyses. Boxes \textbf{1}-\textbf{12} are labelled to match the results in \cite{Funk06}, whilst boxes \textbf{a}-\textbf{z} are new to this analysis. The fitted major axis used for the extent measurements is overlaid and the position of the pulsar (located at the centre of box \textbf{9}) indicated. Significance contours shown correspond to 5, 10, 30, and 50 $\sigma,$ respectively. LS~5039 can be identified as a high-significance point source. Five spectral boxes (\textbf{I}-\textbf{V}) covering the HESS\,J1826--130 region are also shown.}
\label{fig:spectralmap}
\end{figure*}

\section{Spatially resolved spectral map}
\label{sec:spectralmap}

Given the considerable size of the nebula, which is much larger than the H.E.S.S. PSF of $0.064^\circ$ (analysis A), it is appropriate to use the rich amount of information contained in this dataset through more detailed spectral analysis, rather than reducing studies to focus on the pulsar, the peak of the emission, or a primary direction. To this end, a grid of $0.26^{\circ}\times 0.26^{\circ}$ boxes for which the size used in \cite{Funk06} was adopted, was defined based on the 5$\sigma$ significance contour of the nebula emission, from which a spectral analysis was performed for each box. Serving to underline the enormous size of this nebula compared with other similar TeV sources, a total of 38 such boxes were identified, the first 12 of which are defined to coincide with those used in \cite{Funk06}. Because of increased levels of background systematics encountered in the spectra at low energies, which can bias the spectral power-law fit, particularly towards the outer regions of the nebula, a higher energy threshold of $\sim300~$GeV was adopted for this analysis (but otherwise, the same event selection cuts were used as before). 
For the majority of the spectral boxes, the flux was not sufficient for a reliable fit to curved spectral models, but for six of the innermost regions (boxes \textbf{4}-\textbf{9}), curved models were preferred over a power-law fit at the $4\sigma$ level.   In order to avoid a bias in the resulting spectral index, the spectra from each box were fitted with a power law within a limited range of $0.3-5$~TeV.  Fitting a power law to the entire energy range without an upper limit led to a systematic shift of 0.1 in spectral index, which biased the results towards reporting softer indices. Therefore, the fitted range was limited to 5~TeV, which was chosen such that this upper limit falls at an energy below all of the statistically significant cut-off energies. The integrated flux above 1~TeV was found to be consistent within errors of the integrated flux from $1-5$~TeV. 

The results of a power-law fit to each box are shown in figure \ref{fig:spectralmap}, where an additional five boxes covering HESS\,J1826--130 are shown for completeness. Significance contours at 5, 10, 30, and 50 $\sigma$ are shown, using in this case a larger correlation radius of $0.1^\circ$, that is, about half a box width, in order to better correspond to the resolution scale provided by the size of the spectral boxes. 

By eye, a clear trend of softening of the spectral index with increasing distance from PSR\,B1823--13 (located at the centre of box \textbf{9}) is visible. Towards HESS\,J1826--130, the spectral index can be seen to remain comparatively hard with respect to the rest of the nebula, with similar spectral indices across the boundary of the two sources on the resolution scale of the spectral regions defined. Over the HESS\,J1826--130 region (boxes \textbf{I}-\textbf{V}), the level of contamination of HESS\,J1826--130 by HESS\,J1825--137 may vary. (A detailed study of the HESS\,J1826--130 region will be made in a forthcoming publication.) 

Moreover, the spectral index appears to remain harder along the major axis of the nebula. The power-law index does not only soften with distance along the major axis, but also with distance away from the axis on either side.  The spectral properties of the nebula become softer towards the 5 $\sigma$ contour in all directions away from the neighbouring source HESS\,J1826--130. 

\begin{figure}
\begin{center}
\includegraphics[trim=0mm 4mm 2mm 4mm,clip,width=0.9\columnwidth]{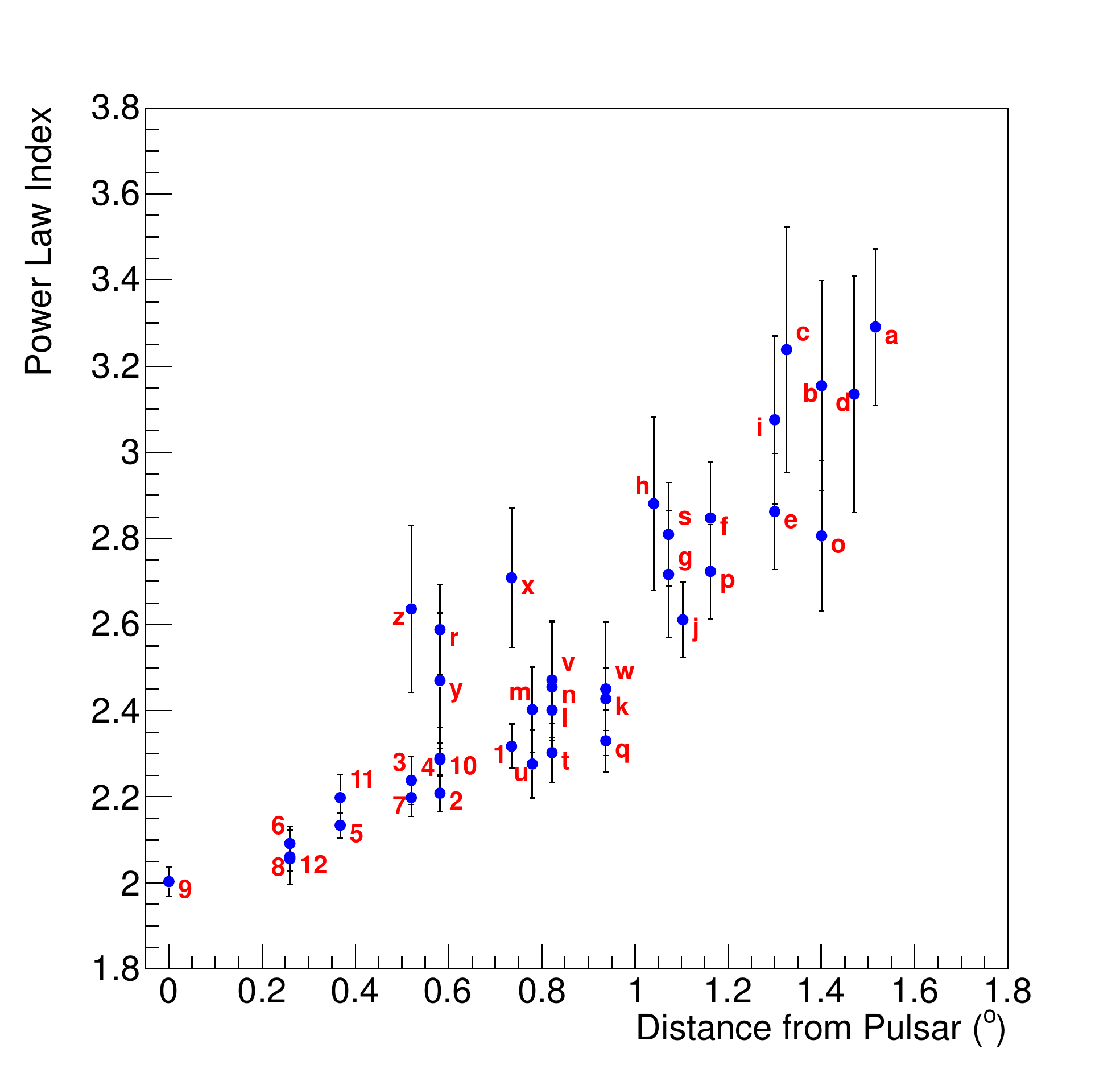}
\includegraphics[trim=0mm 4mm 2mm 4mm,clip,width=0.9\columnwidth]{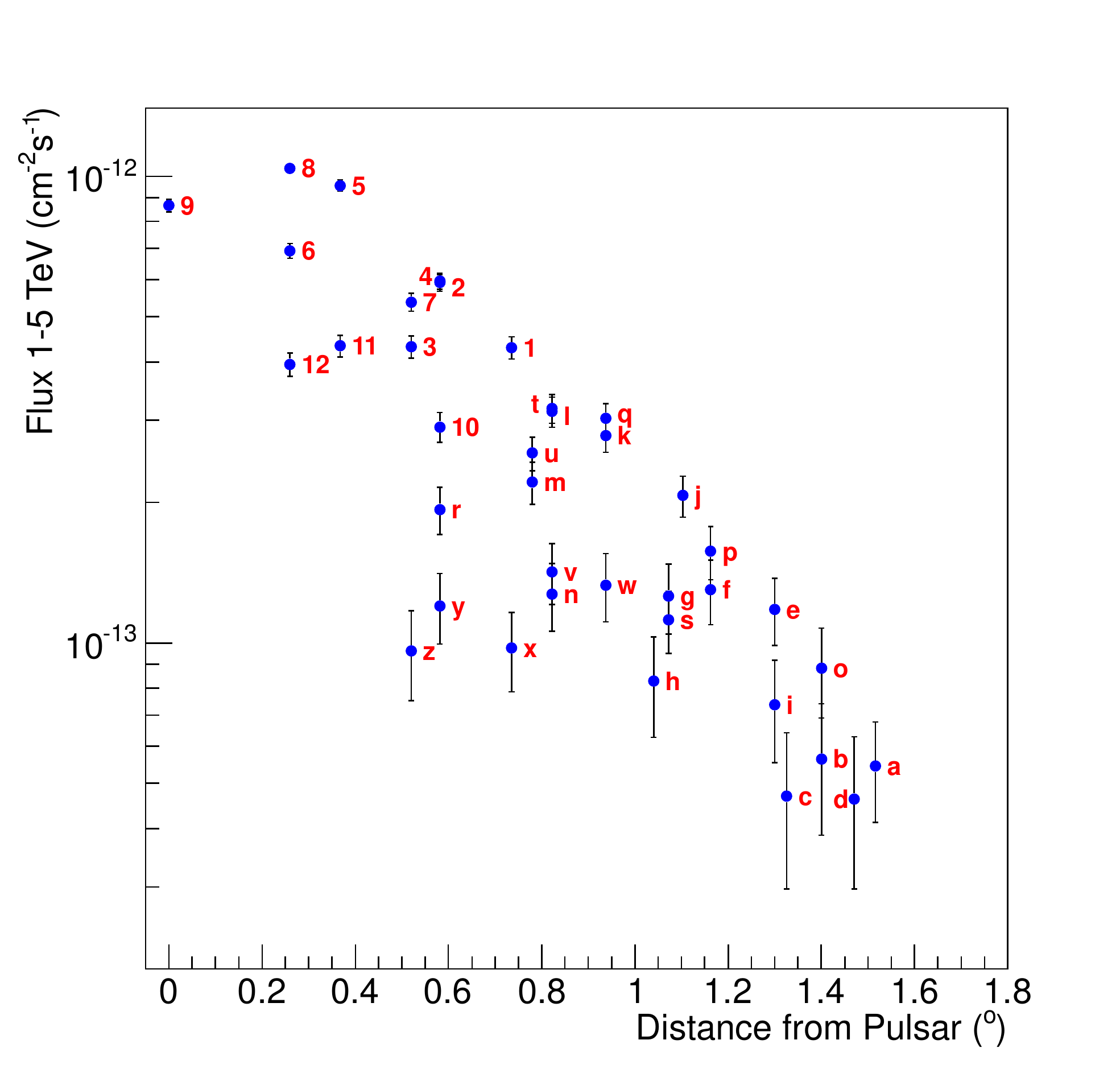}
\includegraphics[trim=0mm 4mm 2mm 4mm,clip,width=0.9\columnwidth]{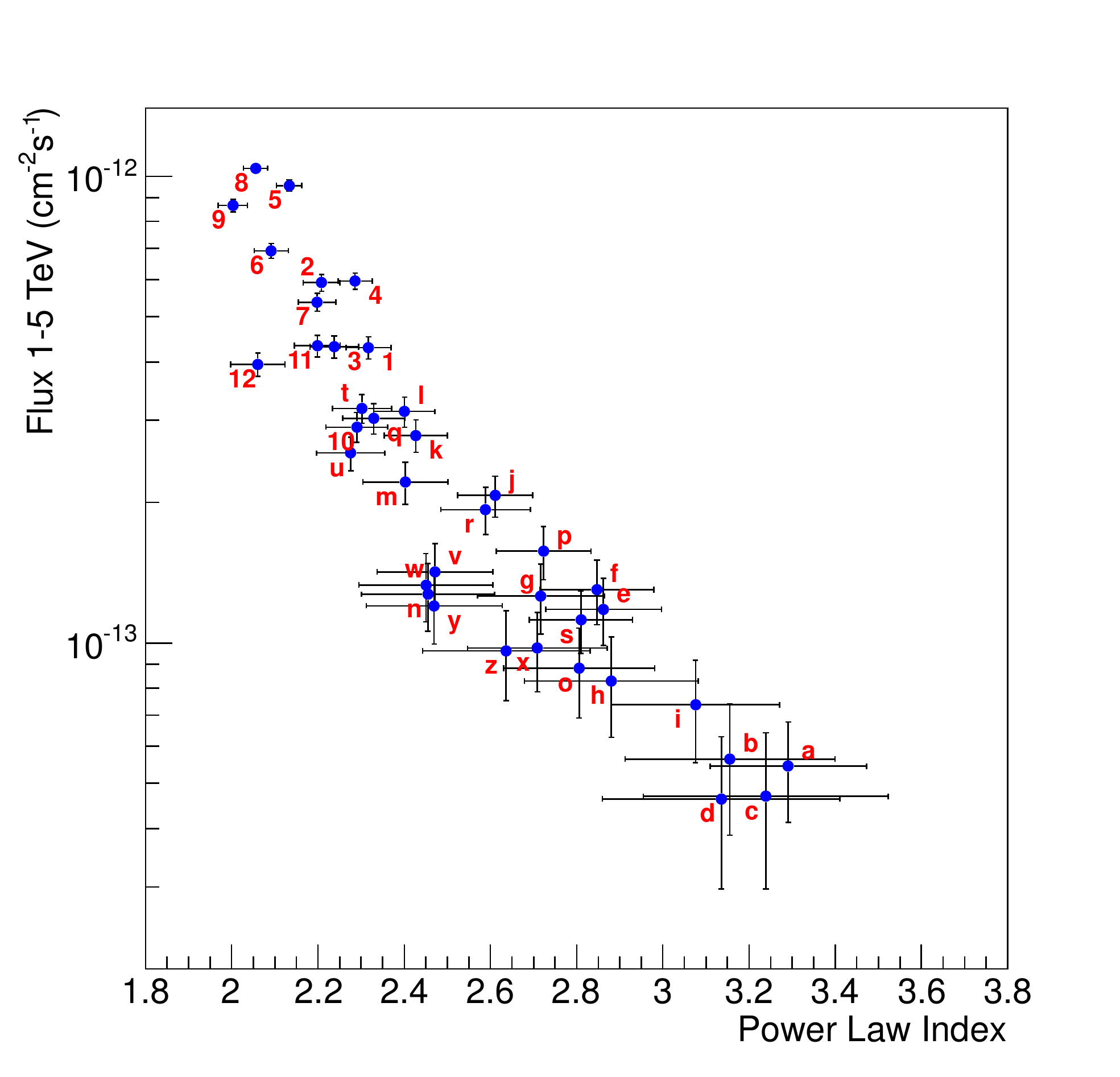}
\caption{Trend of fitted spectral power-law index with distance, flux from $1-5$~TeV with distance, and of flux from $1-5$~TeV with a power-law index from the power-law fits to the spectral boxes depicted in figure \ref{fig:spectralmap}. Correlation coefficients of $0.89 \pm 0.06_{\mathrm{stat.}} \pm 0.17_{\mathrm{sys.}}$, $-0.83 \pm0.03_{\mathrm{stat.}}\pm 0.01_{\mathrm{sys.}}$ , and $-0.92 \pm 0.03_{\mathrm{stat.}}\pm 0.4_{\mathrm{sys.}}$  were found.}
\label{fig:correlations}
\end{center}
\end{figure}

The overall trend of spectral variation throughout the nebula is shown in figure \ref{fig:correlations}, where the variation of spectral index and flux from $1-5$~TeV (from a power-law fit) with distance and with each other is investigated.\ Spectral index and distance are strongly positively correlated, with a correlation coefficient of $0.89 \pm 0.06_{\mathrm{stat.}} \pm 0.17_{\mathrm{sys.}}$. The flux from $1-5$~TeV was found to be anti-correlated with both distance and spectral index, with correlation coefficients of $-0.83 \pm0.03_{\mathrm{stat.}}\pm 0.01_{\mathrm{sys.}}$ and $-0.92 \pm 0.03_{\mathrm{stat.}}\pm 0.4_{\mathrm{sys.}}$ , respectively. (Spectra for each of the 38 spectral boxes are available as part of the supplementary information.)

Although it might be considered to add further regions to this grid, we found that when a $5 \sigma$ significance contour would effectively divide a spectral box into two, a significant spectrum could not be obtained by either the primary analysis or the independent cross-check analysis; such regions were therefore not used in figures \ref{fig:spectralmap} and \ref{fig:correlations}. All of the remaining spectral regions, with the majority of the area lying within the $5 \sigma$ significance contours, have a significance greater than $5 \sigma$ and can therefore be confidently included as contributing to the nebula emission. 
We note, however, that the systematic errors increase particularly towards the outer edges of the nebula (e.g. regions \textbf{a}-\textbf{d} and \textbf{w}-\textbf{z}, see table \ref{tab:spectralboxes}). This corresponds to cases where the significance of the spectral box was close to $5 \sigma$ in either the primary analysis or the independent cross-check analysis and increased background systematics are likely to contribute. 
The 38 $0.26^\circ \times 0.26^\circ$ boxes correspond to a total area of $2.57$ square degrees, which provides a lower limit of the projected area of HESS\,J1825--137 on the sky. 

Of particular interest are boxes \textbf{a} and \textbf{d}; these two regions lie comfortably within the $5\sigma$ significance contour, with spectral significances of 9$\sigma$ and 6 $\sigma,$ respectively. The centres of both regions lie at distances in excess of 100\,pc, as shown in table \ref{tab:spectralboxes} (assuming a 4~kpc distance), the largest distances of a spectral box from the pulsar, closely followed by the 98\,pc distance to regions \textbf{b} and \textbf{o}. This means that HESS\,J1825--137 is one of the largest, if not the largest, $\gamma$-ray PWN(e) currently known in terms of its intrinsic size. This feature merits a dedicated study \citep{Khangulyan17}.

\section{Total nebula flux}
\label{sec:totalflux}

The large size of HESS\,J1825--137, together with its complex morphology and the presence of multiple sources in the surrounding region mean that it is not effective to define (with respect to the pointing position of the telescopes) background extraction regions that are spatially symmetric with respect to the source region.
In order to determine an overall spectrum for the entire nebula, we therefore combined the spectra from the individual boxes defined above. Similar approaches have previously been adopted by H.E.S.S., such as that used in \cite{Eger16}. 
However, in the case of HESS\,J1825--137, we encountered several problems with this approach. For example, many of the regions in the outer reaches of the nebula have soft spectra (power-law index $\gtrsim 2.5$) and only few high-energy events, $E \gtrsim20$~TeV, such that despite the known presence of significant emission above $20$~TeV close to the pulsar, the total spectrum flux over a much larger area is poorly estimated at $E \gtrsim20$~TeV due to the increased statistical noise. To circumvent this, we gradually ceased to combine all boxes into the total spectrum, using a criterion based on the contribution of each region to the total flux, as described in section \ref{sec:combimethod} below. By combining regions throughout the nebula, we obtained the total flux, but we did not perform a single spectral fit to the spectrum.

For strongly energy-dependent sources, a simple spectral fit to the entire nebula that treats the nebula as a single-zone particle population is equivalent to taking the average of the spectral indices throughout the nebula, which may not be particularly informative, as figure \ref{fig:correlations} demonstrates. To this end, we do not interpret the total flux from the nebula in terms of any particular spectral model; rather, we present a total integrated flux of the nebula emission to interpret the overall source strength. 

\begin{figure*}
\centering
\includegraphics[width=1.8\columnwidth]{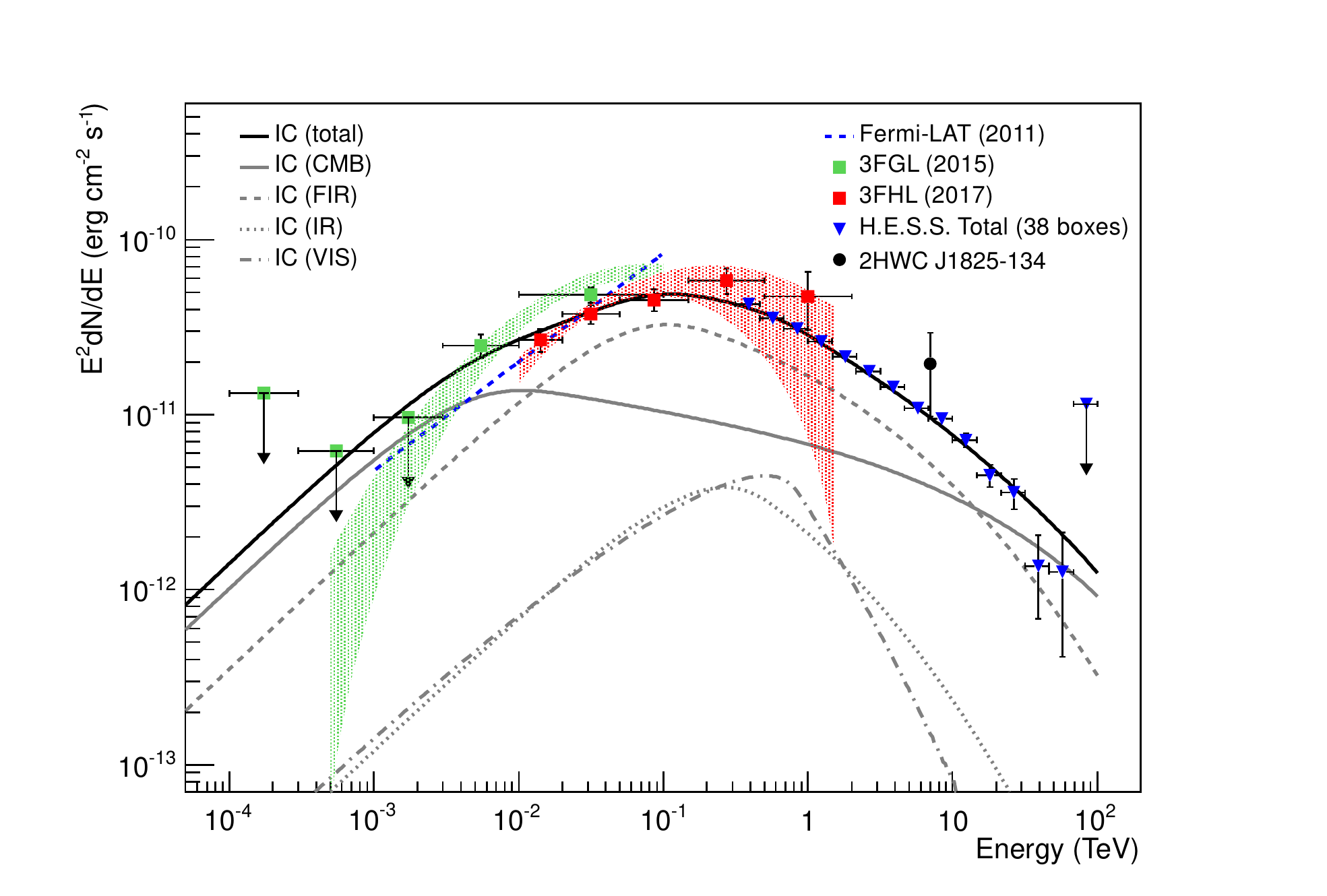} 
\caption{ Spectral energy distribution of HESS\,J1825--137 showing the total combined H.E.S.S. energy flux with statistical error bars, as determined with the method outlined in section \ref{sec:combimethod}. The Fermi-LAT spectral fit from \citep{Grondin11} is also shown, together with spectra from the 3FGL and 3FHL catalogues \citep{3fglAcero15,3fhlAjello17}, both of which were best fit by a log-parabola spectral model. A plausible IC emission model with multiple contributing radiation fields is shown, using the parameterisation of the \cite{Popescu17} radiation model and the Naima modelling package \citep{Zabalza15}. The emission is mostly dominated by the FIR field; the CMB becomes dominant at a few tens of TeV and below $\sim10$\,GeV. } 
\label{fig:totalsed}
\end{figure*}

\subsection{Total nebula spectrum}
\label{sec:totalspectrum}

The resulting spectral flux obtained for the nebula from the summation of the 38 independent $0.26^{\circ}\times0.26^{\circ}$ spectral boxes is shown in figure \ref{fig:totalsed}. For illustration, the HESS\,J1825--137 spectrum obtained from Fermi-LAT \citep{Grondin11} is also shown, although we recall that the region from which this spectrum is obtained, based on a morphological symmetric Gaussian fit template to the Fermi data, does not fully correspond to the same spatial region. The 2011 analysis and the 3FGL (the third LAT Catalogue, 20~MeV to 300~GeV, \citet{3fglAcero15}) both use a Gaussian with $\sigma = 0.56^{\circ}$ as the spatial template, whereas the 3FHL (the third LAT High-Energy Catalogue, 10~GeV to 2~TeV, \cite{3fhlAjello17}) uses a Gaussian with $\sigma = 0.75^{\circ}$, which is consistent with the Fermi Galactic Extended Source catalogue best-fit Gaussian with $\sigma = 0.79 \pm 0.04$ \citep{Ackermann17}. Whilst these templates are only approximations to the source morphology, in figure \ref{fig:totalsed} the IC peak may be inferred to occur at around 200~GeV.

The total flux above 300\,GeV was found to be $(6.07 \pm 0.13_{\mathrm{stat}} \pm 0.07_{\mathrm{sys}}) \times 10^{-11}\mathrm{cm}^{-2}\mathrm{s}^{-1}$, corresponding to $\sim 54\%$ of the flux of the Crab nebula. Above 1~TeV, a flux of $(1.12 \pm 0.03_{\mathrm{stat}} \pm 0.25_{\mathrm{sys}}) \times 10^{-11}\mathrm{cm}^{-2}\mathrm{s}^{-1}$ was obtained, such that the proportion relative to the Crab nebula increases slightly to $\sim 64\%$ of the Crab nebula flux. The source remains strong at $\sim 61\%$ of the Crab nebula flux above 10~TeV. It is impressive that the $\gamma$-ray flux is so strong given that the distance to HESS\,J1825--137 is approximately twice that of the Crab nebula (at 2~kpc), despite (or perhaps because of) the significantly older age of the system. Interestingly, as part of the second HAWC catalogue, it was found that the confused source associated with the combined emission from the region, 2HWC\,J1825--134, is brighter than the Crab nebula at 7~TeV \citep{Abeysekara17}. This point is included in figure \ref{fig:totalsed} and is compatible within the HAWC systematic errors of the total nebula spectrum from HESS\,J1825--137.

The total nebula spectrum could be well described by a simple model of IC scattering including CMB, FIR, IR, and VIS radiation fields in a parameterisation of \cite{Popescu17} using the Naima modelling package \citep{Zabalza15}. An additional UV component was included for a more accurate parameterisation of the Galactic radiation field, but the contribution of this component to the total IC scattering was found to be negligible: it peaks below the scale of figure \ref{fig:totalsed}. The contribution from the FIR was found to dominate most of the energy range, with a transition to CMB dominance at a few tens of TeV. The electron spectrum could be described by a broken power law with low-energy index  $\Gamma_1=1.4\pm0.1$, a high-energy index of $\Gamma_2=3.25\pm0.02,$ and a break energy of $E_b=0.9\pm0.1$\,TeV for a total energy in electrons of $\sim5.5\times10^{48}\,$erg. 
This phenomenological electron spectrum has no high-energy cut-off; the spectral curvature seen at the highest energies can therefore be entirely attributed to Klein-Nishina effects.

\section{Discussion} 
\label{sec:discuss}

HESS\,J1825--137 extends farther than has previously been seen at TeV energies \citep{Funk06} because our analysis benefits from a better sensitivity to large areas of weaker, lower energy emission. 
In agreement with previous findings, the spectral index of the emission increases with increasing distance from the pulsar as a result of the cooling of the electrons over time. This causes the index to become softer and the high-energy flux to decrease.

For the circular regions with radii of $0.8^\circ$ and $0.4^\circ$, curved spectral models are preferred over a pure power-law fit, which indicates the inherent cooling of particles within the nebula and potentially the presence of multiple particle populations of different ages along the line of sight. In the case of a power law with exponential cut-off fit, the cut-off energy provides some indication of the maximum particle energies reached (typically an order of magnitude higher than the $\gamma$-ray energy). A statistically significant spectral cut-off could be explained by a cut-off in the parent particle spectrum that arises from the acceleration and energy-loss mechanisms. Spatial variation with energy, or equivalently, spectral variation with position within the nebula, indicates that the nebula emission detected from Earth is the summation of the contributions from multiple particle populations with different spectral properties. Although the transitions between such populations may be smooth, \cite{VanEtten11} showed that a multi-zone modelling approach is more successful in reproducing the properties of the present-day nebula. 

The major axis to the emission from the nebula was found to lie at $208^\circ$ counter-clockwise from north, almost perpendicular to ($72^\circ$ away from) the projected trajectory of the pulsar across the sky from its extrapolated birth position. 
If the direction of bulk particle motion were aligned with the particle wind, it would be expected that the emission were aligned with (but be opposite to) the proper motion in the case of young PWNe \citep{Kaplan08}. 
Therefore, the non-alignment provides a strong indication that additional effects influence the evolution, such as the conditions of the ambient medium. The steep drop observed towards the north might be due to a combination of physical effects (such as asymmetric crushing by a reverse shock from the progenitor supernova) and projection effects, especially as the component of the pulsar velocity along the line of sight is not known.

A method of characterising the nebula extent was established and found the characteristic extent ($r_{1/e}$) a factor of $1.6\pm 0.1$ larger towards the south than on the northern side of the nebula. Although the northern emission was also seen to decrease towards the highest energies, the variation in extent with energy was found to be less pronounced than in the southern direction and is compatible with no significant change. 

Considering the changing nebula extent with energy, we fit by assuming a relationship of $E_\gamma \propto E_e^2$ as for the Thomson regime and found that a purely diffusive particle transport model was disfavoured, with some significant contribution from advective transport mechanisms likely. 
In this case, the motion can be described as a radial convective flow with diffusion occurring in directions perpendicular to this flow. We find that the bulk flow can be described by a velocity profile $v(r) \propto r^{-0.7\pm 0.2_{\mathrm{stat}} \pm 0.3_{\mathrm{sys}}}$, under the assumption of constant flow density, although a variable density profile is not excluded. 

This finding is in agreement with several previous models of this source. The suggestion that the nebula is particle dominated, with a particle pressure that is still significant against the surrounding ISM pressure, was made by \cite{deJager09}. This would imply that advective processes may still be playing a significant role in the nebula morphology.
The velocity profile obtained from the H.E.S.S. data is in good agreement with the $v(r) \propto r^{-0.5}$ velocity profile arising from the model of \cite{VanEtten11}, whose best-fit model incorporated both advection and diffusive processes to describe the transport. %

In section \ref{sec:extentransport} we considered the Thomson regime dependency of $E_\gamma \propto E_e^2$ and the pure KN regime relation $E_\gamma\sim E_e$. Although IC scattering from the CMB radiation field is well within the Thomson regime, the total energy density also comprises significant contributions from dust (FIR), IR, and starlight (VIS) radiation fields. The dominant contribution to the IC scattering comes from the FIR field, for which KN effects are not negligible.
Simulations using the EDGE (\cite{Lopez17ICRC,Hahn15} modelling package indicate that including KN effects and a realistic model of the radiation fields leads to energy-dependent variations in the relation between $E_e$ and $E_\gamma$. For the results shown in figure \ref{fig:REmodellines}, a magnetic field strength of $5\,\mu$G was adopted; for the observed reduction in nebula size at lower energies, a magnetic field strength of $\sim12\,\mu$G would be necessary, which is not supported by X-ray measurements \citep{Uchiyama09}, however. Fully accounting for all effects, that is, deriving $\gamma$-ray spectra from the full spatial and spectral convolution of electron spectra and the exact radiation cross-sections, as well as including the line-of-sight integration, would require more extensive modelling that is beyond the scope of this paper.

Throughout, we have assumed a characteristic age of 21~kyr, but if the pulsar has a braking index lower than 3, as suggested by several models, the estimated age of the nebula would increase. One example would be to decrease the braking index of the nebula to $n=2$ or lower to account for the large size of the nebula, thereby doubling the estimated age to $\sim 40$\,kyr, as suggested by \cite{VanEtten11}. This would correspondingly alter the timescale for cooling losses, and the lower bound of the fit of the extent as a function of energy would be halved. This would introduce one extra point into the fit, and would change the fitted slope to become slightly less steep, which would yield a velocity profile for advection closer to $v(r)\propto r^{-1}$ and a diffusion index of $\delta \sim 0$. Following this scenario, the slope lies on the edge of the overlap region that is compatible with diffusion and advection as dominant transport processes. Whilst in this case energy-independent diffusion cannot be ruled out, it remains likely that advective processes would continue to play a major role.

The nebula appears to shrink towards lower energies (below 700~GeV), although at around 200~GeV, measurements approach the H.E.S.S. energy threshold and background systematics become more influential, which is reflected by the increased uncertainties at these lowest energies. Analysis B is more sensitive in this region because CT5 is included. This yields an additional point at energies below $\sim$125\,GeV. 

At energies below 700~GeV, the lifetime of the particles against cooling losses is longer than the 21~kyr age of the nebula, such that the size may be expected to increase with increasing energy, with a maximum size occurring at $\sim$700~GeV \citep{AAV95}. %
Energy-dependent extent measurements that cover energies below 700~GeV and extend below the H.E.S.S. energy range by complementary facilities would provide further insights into the dominant processes affecting HESS\,J1825--137, although no dramatic energy dependence of the morphology within this energy range was reported as part of the Fermi-LAT results \citep{Grondin11}.

The overall extent of the nebula was found to reach an intrinsic diameter of $\sim100$~pc (assuming a 4~kpc distance) as measured to the outermost spectral boxes, rendering HESS\,J1825--137 an exceptionally large PWN \citep{Abdalla17pwnpop}. The H$\alpha$ ridge identified in \cite{Voisin16} and \cite{Stupar08} implies a supernova size of $\sim120$~pc radius.
\cite{Khangulyan17} offer an explanation for the anomalously large extent of the nebula through significant energy injection by the pulsar at early times, such that the pulsar contribution dominates the kinetic energy of the supernova ejecta, invoking an initial spin-down period $P_0\sim1$~ms with a surrounding density of $n\gtrsim1\mathrm{cm}^{-3}$ rather than the alternative of requiring an extremely low-density ambient medium a priori ($n \leq 10^{-2}\mathrm{cm}^{-3}$) in the case of the Sedov solution. 
The PWN expansion occurs preferentially into low-density regions, which is supported by the observations of multiple molecular clouds in the region, away from the peak of the TeV emission.

This scenario is supported by the best-fit model of \cite{VanEtten11}, which incorporates a velocity profile, magnetic field profile, and diffusion and adopts an initial spin period of $P_0 \sim 13$~ms with a braking index of $n=1.9$ to provide adequate energy to power the nebula. In all cases, although advection is responsible for bulk flow within the nebula, a significant diffusive contribution is still expected, especially towards the outer reaches of the nebula and in order to account for the nebula breadth, rather than only the extent along the major axis. A diffusive scenario is compatible with our results; a 100\,pc extent that is achieved within an approximately 21\,kyr lifetime implies a diffusion coefficient of $D_0\sim1.4\times 10^{29}\,\mathrm{cm}^2\mathrm{s}^{-1}$, which is within the expected range for ISM diffusion at 1\,TeV \citep{Maurin01}. The flow velocity implied by this nebula size and assumed age is typically $\sim\,0.01$c.

The spatially resolved spectral map of the nebula further illustrates the variation not only as a function of distance from the pulsar, but also more generally as a function of location within the nebula. 
The (anti-)correlation between the fitted spectral parameters (power-law index, flux from 1-5~TeV) and distance from the pulsar is high. However, the distance is always the projected distance; therefore the spectral properties are given as obtained from the superposition of emission along the line of sight. It is expected that multiple particle populations with differing ages and spectral properties are present in each region. 

The total spectrum reduces the information that is available about the nebula to the total flux. Within the HGPS, it was found that the morphological description of the nebula was improved by using three Gaussian components rather than a single Gaussian to model the total VHE $\gamma$-ray emission, which is another indication that the nebula composition is very complex \citep{Abdalla18hgps}. As figure \ref{fig:correlations} suggests, new insights are more likely to be gained by approaches that incorporate multiple particle populations, such as that of \cite{VanEtten11}.

The occurrence of the IC peak in figure \ref{fig:totalsed} at around 100~GeV would correspond to a cooling of the electron spectrum at particle energies of around 1~TeV, in agreement with the IC model we showed. %
The contribution from the near-IR and optical photon fields (VIS) through background starlight to the total IC scattering was verified by a simple model to be negligible and KN suppressed \citep{Zabalza15}. The effects of KN losses on the total spectrum  (figure \ref{fig:totalsed}) were found to be small, with a total energy in electrons of $\sim5.5\times10^{48}\,$erg.
The total nebula $\gamma$-ray spectrum exhibits some spectral curvature at $\sim$10\,TeV and can be well described by a broken power-law electron spectrum without a spectral cut-off. This indicates that this curvature can be fully accounted for by KN effects.

The total $\gamma$-ray luminosity above 300~GeV that we obtained from the spectrum shown in figure \ref{fig:totalsed} is $L_\gamma = 3.1\times 10^{35}\mathrm{erg\,s}^{-1}$ assuming a 4\,kpc distance. For the nebula volume, it was assumed that the emission corresponding to each spectral box occupies a line-of-sight depth equivalent to $\sqrt{(r_{\mathrm{max}}^2 - r_{\mathrm{box}}^2)}$, where $r_{\mathrm{max}}$ is the distance to the outermost region and $r_{\mathrm{box}}$ is the distance of the centre of that box from the pulsar, such that the total volume is given by the sum of the projected box areas multiplied by the assumed depths. The total energy in electrons and total nebula volume yields an energy density throughout the whole nebula of $\sim 0.1\,\mathrm{eV\,cm}^{-3}$.

For comparison, the innermost region of the nebula, box \textbf{9} in figure \ref{fig:spectralmap} centred on the pulsar, has a $\gamma$-ray luminosity above the threshold of $L_\gamma = 2.0\times 10^{34}\mathrm{erg\,s}^{-1}$. A total energy in electrons of $1.0^{+0.8}_{-0.2} \times 10^{47} \mathrm{erg}$ is obtained for box \textbf{9} using the same radiation field model as shown in figure \ref{fig:totalsed} and fitting a broken power-law electron spectrum with the low index fixed to $\Gamma_1=1.4$, that is, to the nebula value. As an edge case, we consider box \textbf{a} (at 106\,pc from the pulsar) with a $\gamma$-ray luminosity above threshold of $L_\gamma \approx 6\times 10^{33}\mathrm{erg\,s}^{-1}$ and total energy in electrons of $5^{+3}_{-2}\times 10^{47} \mathrm{erg}$. 
With fixed $\Gamma_1$, it was found that the high index $\Gamma_2$ was considerably softer and break energy lower for box {\bf a} than for box {\bf 9}, 
as expected for an electron population that cools with age and increasing distance from the pulsar. 

This approximation for the line-of-sight depth to obtain the corresponding box volumes yields energy densities of $\sim\,0.06\,\mathrm{eV\,cm}^{-3}$ and $\sim\,1\,\mathrm{eV\,cm}^{-3}$ for boxes {\bf 9} and {\bf a,} respectively, a consequence of the assumption that the nebula expansion along the line-of-sight depth is far greater closer to the pulsar than in the outer reaches of the nebula. Nevertheless, this is only an approximation to the average value for each box. The true energy density is likely to vary within the nebula along the line of sight.

HESS\,J1825--137 serves as the archetypal example of a pulsar wind nebula with strong energy-dependent morphological properties, but it remains one of only a few known examples. Whether this is due purely to a fortuitous observation angle and to the age of the system, or if it is also influenced by the chance location and corresponding ambient conditions remains to be determined. Certainly the evolution and history of the nebula play a significant role in determining the properties that are currently observed.

As demonstrated here, simple assumptions already lead to insights into the dominant particle transport processes. Simple particle population models including $\gamma$-ray data in conjunction with X-ray data can also lead to insights into the variation in magnetic field and particle energy density throughout the nebula, although caveats are associated with this approach. 
The physical extent of the nebula along the line of sight may be greater for the $\gamma$-ray nebula (less energetic electrons that undergo IC scattering) than for the X-ray nebula (more energetic electrons produce synchrotron radiation). When we assume that the physical extent of the nebula is the same for the X-ray and $\gamma$-ray data, the resulting magnetic field strengths are likely to be somewhat lower than expected, as an average field strength over the larger volume (occupied by the less energetic electrons that produce $\gamma$-ray emission) is obtained. Multi-zone modelling approaches may be able to overcome these difficulties by incorporating multiple zones that are superimposed along the line of sight.

The reported centroid of the HAWC emission 2HWC\,J1825--134 is closer to the previously reported location of HESS\,J1826--130 than to HESS\,J1825--137. This leads to an association by HAWC with the former source. However, in practice it is expected that the centroid of emission from HESS\,J1825--137 would appear farther north (towards the pulsar) than the H.E.S.S. peak because i) the hardest spectra reported from H.E.S.S. for the source are those closest to the pulsar, and ii) the energy threshold of HAWC is higher ($\sim1-10$\,TeV) \citep{Abeysekara17}.
By comparing our results to the HAWC catalogue, we find that the sum of the flux from the three H.E.S.S. sources in the region amounts to$ \text{ about two-thirds}$ of the flux from the extended analysis of 2HWC\,J1825--134 (reported for a $0.9^\circ$ region at 7~TeV) shown in figure \ref{fig:totalsed}, although the total flux from HESS\,J1825--137 alone is still compatible within the 50\,\% systematic error on the flux reported by HAWC \citep{deNaurois06,Ang1826ICRC17}.

Future observations of HESS\,J1825--137 with the forthcoming Cherenkov Telescope Array (CTA) could further our understanding of the physical processes within PWNe \citep{Acharya13,deOnaWilhelmi13}. In particular, the increased field of view will help to reduce background systematics that currently affect the outer regions, whilst the improved angular resolution will help to distinguish the multiple sources in the region and enable further investigations of the spectral variation throughout the nebula, which is hinted at in the comparative spread of flux points in figure \ref{fig:correlations}. Additionally, the wider energy range that becomes available with three telescope sizes will enable exploration of the extent of diffuse low-energy emission and the statistics of the innermost high-energy emission. With increased energy resolution, we can look forward to the unambiguous detection with CTA of energy-dependent morphology in multiple PWNe.
As HESS\,J1825--137 is also one of the strongest Galactic TeV sources, the first data on the nebula with CTA may be anticipated as part of the first observations of the Galactic plane. Undoubtedly, this source still has much to offer as the TeV $\gamma$-ray field continues to make technical advancements. 

\section*{Acknowledgements}
\label{sec:acknowledge}

The support of the Namibian authorities and of the University of Namibia in facilitating 
the construction and operation of H.E.S.S. is gratefully acknowledged, as is the support 
by the German Ministry for Education and Research (BMBF), the Max Planck Society, the 
German Research Foundation (DFG), the Helmholtz Association, the Alexander von Humboldt Foundation, 
the French Ministry of Higher Education, Research and Innovation, the Centre National de la 
Recherche Scientifique (CNRS/IN2P3 and CNRS/INSU), the Commissariat \`{a} l'\'{e}nergie atomique 
et aux \'{e}nergies alternatives (CEA), the U.K. Science and Technology Facilities Council (STFC), 
the Knut and Alice Wallenberg Foundation, the National Science Centre, Poland grant no. 2016/22/M/ST9/00382, 
the South African Department of Science and Technology and National Research Foundation, the 
University of Namibia, the National Commission on Research, Science \& Technology of Namibia (NCRST), 
the Austrian Federal Ministry of Education, Science and Research and the Austrian Science Fund (FWF), 
the Australian Research Council (ARC), the Japan Society for the Promotion of Science and by the 
University of Amsterdam. We appreciate the excellent work of the technical support staff in Berlin, 
Zeuthen, Heidelberg, Palaiseau, Paris, Saclay, T\"{u}bingen and in Namibia in the construction and 
operation of the equipment. This work benefited from services provided by the H.E.S.S. 
Virtual Organisation, supported by the national resource providers of the EGI Federation.

\bibliographystyle{aa}
\bibliography{J1825references}

\section*{Appendix}

\subsection*{Method of combining spectra}
\label{sec:combimethod}

In order to determine the total spectrum, the flux was combined from the 38 box-shaped spectral extraction regions as follows. The independent spectrum for each box was determined first using fixed energy binning.

The total flux $\Phi_n$ for each energy bin $n$ was found by summing the flux over all boxes $i$, determined as follows. The flux expectation, $\phi_{i,n}$, is calculated from the fitted power law for each box:

\begin{equation}
\phi_{i,n} = I_{0,i} \times \left(E_{\mathrm{min},n}+\frac{\Delta E_n}{2}\right)^{-\Gamma_i}~,
\label{eq:fluxexp}
\end{equation}

\noindent where $I_{0,i}$ is the flux normalisation at 1~TeV for box $i$, $E_{\mathrm{min},n}$ the minimum energy of the bin $n$ and $\Gamma_i$ the fitted power-law index. 

The expected flux was normalised by the ratio of the excess counts measured ($\mathrm{Excess}_{i,n}$) to the expected number of excess counts ($\mathrm{ExpExcess}_{i,n}$), given by

\begin{equation}
\mathrm{ExpExcess}_{i,n} = \phi_{i,n} A_{i,n} t_{i,n} \Delta E_{i,n}~,
\end{equation}

\noindent for an effective area $A_i$, exposure time $t_i$ , and bin width $\mathrm{d}E_i$ for each energy bin $n$ and each box $i$. These parameters are given as a result of the spectral analysis of each box. 

The error on the total flux was obtained by combining the statistical uncertainties that are due to the number of ON and OFF events ($\mathlarger{\sigma} (\mathrm{N_{ON,i,n}})$ and $\mathlarger{\sigma} (\mathrm{N_{OFF,i,n}})$) separately, assuming that $\mathrm{N_{ON}}$ and $\mathrm{N_{OFF}}$ follow a normal distribution. Errors on the number of counts in individual bins were estimated as $\sigma(\mathrm{N_{ONi,n}}) = \sqrt{(\mathrm{N_{ON,i,n}})}$, and analogously for the $\mathrm{N_{OFF}}$.

Assuming that the ON errors are fully uncorrelated between spectral boxes, these error terms are summed in quadrature for each energy bin over all boxes, such that $\mathlarger{\sigma}^2 (\mathrm{N_{ON,n}}) = \sum_i  \mathlarger{\sigma}^2 (\mathrm{N_{ON,i,n}}) $.
However, given the number and proximity of the box regions, it is certainly possible that some OFF events used for background estimation from non-excluded regions are additionally used in the spectral analyses of nearby boxes. 
The level of overall correlation of OFF events between all boxes was found by constructing a correlation matrix with elements filled by the fraction of unique OFF events between each pair of spectral boxes. The fraction of shared events between each pair was found based on the shared time-stamps of events. 
This led to an additional term being added to the uncorrelated error for each energy bin taking this correlation matrix into account, such that the total OFF error is then given by

\begin{equation}
\mathlarger{\sigma}^2 \mathsmaller{(\mathrm{N_{OFF,n}})} = \sum_i \mathlarger{\sigma}^2 \mathsmaller{(\mathrm{N_{OFF,i,n}})} + \sum_{i}\sum_{j,j \neq i} \rho_{i,j} \mathlarger{\sigma} \mathsmaller{(\mathrm{N_{OFF,i,n}})} \mathlarger{\sigma} \mathsmaller{(\mathrm{N_{OFF,j,n}})}~,
\end{equation}

\noindent where $\rho_{i,j}$ is the OFF event correlation matrix element between boxes $i$ and $j$.

Finally, these contributions were combined to give an error for each flux point of

\begin{equation}
\mathlarger{\sigma}(\Phi_n)^2 = \left(\frac{\Phi_n}{\sum_i \mathrm{Excess}_{i,n}}\right)^{2} \times \left(\mathlarger{\sigma}^2 \mathsmaller{(\mathrm{N_{ON,n}})} + \bar{\alpha_n}^2\mathlarger{\sigma}^2\mathsmaller{(\mathrm{N_{OFF,n}})}\right)~,
\label{eq:fluxerr}
\end{equation}

\noindent where $\bar{\alpha_n}$ is the weighted average $\alpha_n$ over all boxes; $\bar{\alpha_n} = \sum_i \alpha_{i,n} \mathrm{N_{OFF,i,n}} / \sum_i \mathrm{N_{OFF,i,n}}$ and $\alpha_{i,n}$ is the ratio of the ON exposure to OFF exposure for each box $i$ and energy bin $n$.

In forming the total spectrum, the changing size of the nebula with energy needs to be accounted for in a robust way to avoid adding multiple non-significant boxes at the highest energies, which can lead to greater statistical uncertainties in the flux. 
 To this end, the flux from a given box region was not included when the relative contribution dropped below 1\% of the total of the fitted power-law models. Boxes for which this condition occurred were omitted at all energies above the lowest energy at which the contribution was found to be lower than 1\%.

\end{document}